\documentclass{article}
\usepackage[a4paper, total={6.5in,9in}]{geometry}
\usepackage[utf8]{inputenc}
\usepackage{amsmath}
\usepackage{amssymb}
\usepackage{graphicx}
\usepackage{authblk}
\usepackage{url}
\usepackage[authoryear]{natbib}
\usepackage{adjustbox}
\usepackage{rotating}
\usepackage{comment}
\usepackage{longtable}
\usepackage{xcolor}
\usepackage{ulem}
\usepackage[english]{babel}
\usepackage{setspace}
\usepackage{multirow}
\usepackage{amsfonts}
\usepackage{array} 
\usepackage{flushend} 
\usepackage{stfloats} 
\usepackage{color}
\usepackage{chngpage} 
\usepackage{totcount} 
\usepackage{hyperref}
\usepackage{fix-cm} 
\usepackage{algorithm}
\usepackage{algorithmicx} 
\usepackage{algpseudocode} 
\usepackage{listings} 
\usepackage{crop}
\usepackage{subfloat} 
\usepackage{subfig}
\usepackage{tikz} 
\usepackage{hyperref} 
\usepackage{footnote}
\usepackage{mathrsfs} 
\usepackage{wrapfig} 
\usepackage{amsthm} \usepackage[most]{tcolorbox}
\usepackage{listings}  % for code formatting
\tcbuselibrary{listings, breakable}

\newtcolorbox[auto counter, number within=section]{mybox}[2][]{%
	enhanced,
	colback=blue!5!white,
	colframe=blue!75!black,
	fonttitle=\bfseries,
	coltitle=white,
	title=Box~\thetcbcounter: #2,
	listing only,
	listing options={language=R},
	unbreakable,
	boxsep=1pt,
	left=2pt,      % Set slightly above 0 to avoid clipping border
	right=2pt,
	top=2pt,
	bottom=2pt,
	sharp corners,
	#1
}

\begin{document}
	
	\vspace{-40pt}
	\title{An Introduction to Topological Data Analysis Ball Mapper in R}
	\vspace{-30pt}

	\author[1]{Simon Rudkin \thanks{Full Address: Department of Social Statistics, School of Social Sciences, University of Manchester, Oxford Road, Manchester, M13 9PL, United Kingdom. Email:simon.rudkin@manchester.ac.uk}}
	\affil[1]{School of Social Sciences, University of Manchester, United Kingdom}
	
%	\vspace{-30pt}
%	\author[2]{Wanling Rudkin \thanks{\textbf{Corresponding Author} Full Address: University of Exeter Business School, University of Exeter, Streatham Court, Rennes Drive, Exeter, EX4 4PU, United Kingdom. Tel: +44 (0)7955 109334  Email: w.rudkin@exeter.ac.uk}}
%	\affil[2]{University of Exeter Business School, University of Exeter, United Kingdom}
	\vspace{-40pt}
	\vspace{-20pt}
	
	\doublespacing
	
	\date{}

	\maketitle
	
	\begin{abstract}
		The Topological Data Analysis Ball Mapper (TDABM) algorithm of \cite{dlotko2019ball} provides a model free means to visualize multi-dimensional data. The visualizations are abstract two-dimensional representations of covers of the dataset. To construct a TDABM plot, each variable in the dataset should be ordinal and suitable for representing as an axis of a scatter plot. The graphs produced by TDABM provide a map of the dataset on which outcomes may be charted, models assessed and new models formed. The benefits of TDABM are powering a growing literature. This document provides a step-by-step introduction to the algorithm with code in R.
	\end{abstract}


\section{Introduction}

Consider a dataset $X$ formed of $N$ observations on $K$ variables. The dataset is also equipped with a further variable, $Y$, which has a numeric value for all $N$ data points. Hence point $i$, $i \in \lbrace 1,N \rbrace$ the values of the $K$ variables are $x_{ik}$, $k \in \lbrace 1,K \rbrace$. Data points can then be plotted as a point cloud where each axis of the point cloud represents one of the $K$ variables. The location of a point in the $K$-dimensional space is defined by the values of $x_{ik}$. Topological Data Analysis (TDA) is based upon measures of the point cloud. Where the point cloud has more than two dimensions, $K>2$, the visualization of the point cloud becomes difficult. This short guide explains the use of the Topological Data Analysis Ball Mapper (TDABM) algorithm of \cite{dlotko2019R} using the \textit{BallMapper} package in R\footnote{Throughout reference is made to the R computing language \citep{rbase}. The code accompanying this guide is available at \href{https://github.com/srudkin12/BM-Guide}{https://github.com/srudkin12/BM-Guide}.}.

Sources of data for the point cloud are many. The variables may be drawn as multiple regional aggregates on the same topic as in \cite{rudkin2023economic} study of voting patterns in the United Kingdom referendum on European Union membership. \cite{otway2024shape} uses a similar regional aggregate approach to study voting behaviours in the UK general elections. Alternatively, all variables may be different topics as in \cite{rudkin2024topology} study of the digital divide at the dawn of the Covid-19 pandemic. The $K$ variables may be lags of the same time series, as seen in \cite{rudkin2024return} plotting of the return trajectories of Bitcoin or \cite{rudkin2023regional} study of regional economic performance trajectories. Applications in finance include the consideration of credit risk across a space of financial ratios \citep{qiu2020refining}, and stock returns across a joint distribution of firm characteristics \citep{dlotko2021financial}.

The rationale for mapping the multidimensional space in a single plot is provided by the same arguments on data visualization as made by \cite{anscombe1973graphs} and \cite{matejka2017same}. Whether the aim is to develop an understanding of the structure of the data that is not given in the first and second moments, or understanding statistical models, the ability to visualize data is core. Although this guide does not provide the specific tools to evaluate models, the ability to map is an important first step. Readers are referred to the relevant methodological papers for more. See for example, \cite{dlotko2021financial} and \cite{rudkin2024return} for evaluation of model fit, and \cite{rudkin2024return} for a naive forecaster based on TDABM. \cite{charmpi2023topological} provide an example of a TDABM classifier for credit risk in a similar fashion. A useful discussion on the choice of parameters in TDABM is provided in \cite{dlotko2022topological}. 

There are many alternative means for visualizing multivariate data. Alternatives include the original mapper algorithm of \cite{singh2007topological}, which is also topologically inspired. The original mapper requires a lens function, clustering algorithm and selection of number of clusters. A research agenda is exploring the choices for mapper, but there is still no definitive means for selecting the choice elements. The mapper graph can be very unstable and hence the single choice parameter in TDABM is preferable. Other methods such as Principal Components Analysis or T-SNE require the application of dimension reduction. Because TDABM does not generate data loss, TDABM provides a preferable means to visualize the full structure of the data. This guide focuses on TDABM and does not apply the alternative methods to the example data.

The remainder of the guide is organised as follows. Section \ref{sec:data} introduces the artificial dataset used in this guide. Section \ref{sec:method} introduces the method from a theoretical standpoint. Section \ref{sec:art}  provides an explanation of the way in which the TDABM algorithm is implemented in R. Section \ref{sec:further} discusses the next steps and concludes.

\section{Data}
\label{sec:data}

\begin{figure}
	\begin{center}
		\caption{Raw Dataset}
		\label{fig:raw}
		\includegraphics[width=12cm]{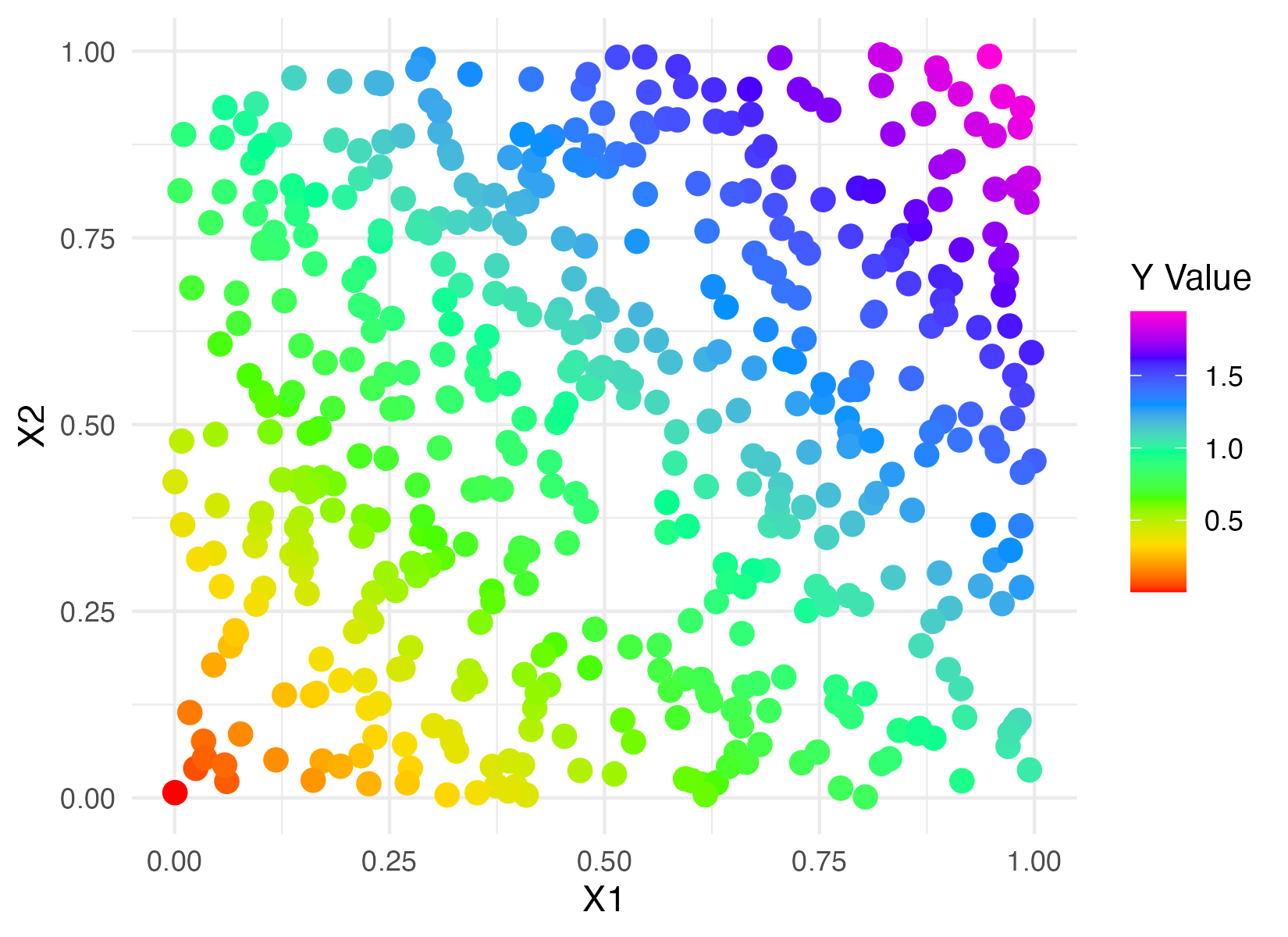}
	\end{center}
	\raggedright
	\footnotesize{Notes: Dataset has $N=500$ observations with $K=2$ variables. Each variable is drawn at random from a uniform distribution such that $X_1 \sim U[0,1]$ and $X_2 \sim U[0,1]$. The outcome variable has $y_i = x_{i1}+x_{i2}$. Points are colored according to $Y$.}
\end{figure}

This note will use the same dataset for the demonstration of the TDABM algorithm. The dataset has $N=500$ data points, with coordinates in $K=2$ dimensions. The variables are denoted as $X_1$ and $X_2$. Both variables are drawn independently form a uniform distribution on $\left[0,1\right]$, $X_1 \sim U[0,1]$ and $X_2 \sim U[0,1]$. For each data point there is an associated outcome variable $Y$ where $y_i = x_{i1}+x_{i2}$. The resulting dataset is plotted in Figure \ref{fig:raw}.

The coloration has a natural pattern with highest values occurring in the North East of the plot. Meanwhile the lowest values are observed to the South West. The 2-dimensional nature of the dataset means that we can see the full pattern of the data without requiring any advanced methods. In reality, there is limited motivation to use a visualisation algorithm like TDABM to represent a 2-dimensional dataset. However, the ability to plot the $K=2$ case on a simple scatter plot allows the opportunity to demonstrate the algorithm clearly. 

The decision to use the uniform distribution with the same minimum and maximum values as the axes means that there is no need to rescale the data prior to running the TDABM algorithm. Like many Machine Learning methods, TDABM is a distance based approach to analyzing data. Where variables exist on different scales then normalisation must be performed prior to running the TDABM algorithm on the data.

The code which is used to generate this dataset in R is provide in Box \ref{box:rcode}. The code demonstrates the simplicity of the construction of the dataset for this example document. Inclusion of Box \ref{box:rcode} removes ambiguity about the data in the discussion that follows. The setting of the random seed is very important to ensure that reproduction of the examples in this guide.

\begin{mybox}[label=box:rcode]{R Code for Dataset}
	This code will produce a bivariate dataset with two uniformly distributed variables and 500 observations:
	\begin{lstlisting}[language=R]
	set.seed(123)  # For reproducibility
	x1<-runif(500,0,1)
	x2<-runif(500,0,1)
	\end{lstlisting}
	We bind the two variables together to produce a \texttt{data.frame} object:
	\begin{lstlisting}
	df1<-as.data.frame(cbind(x1,x2))
	names(df1)<-c("X1","X2")
	\end{lstlisting}
\end{mybox}

\section{Methodology}
\label{sec:method}

The TDABM algorithm begins with a dataset $X$ with $K$ variables. The data is converted into a point cloud in which the location of each data point $i$ is determined by the value of $x_{ik}$ on axis $k$, $k \in \lbrace 1,K \rbrace$. The point cloud for the example dataset is represented in Figure \ref{fig:raw}. The single choice parameter for the TDABM algorithm is the radius of ball to use, $\epsilon$. There is currently no algorithm with which to choose the optimal radius. As with maps of geographic space, there are times when the interest is in local structure, and cases where the interest is in the global structure of the data. The TDABM equivalent is to use a small $\epsilon$ to see local structures and a large $\epsilon$ to see the overall structure of the data. It is recommended that users consider many radii when looking at their data.

\begin{figure}
	\begin{center}
		\caption{Raw Dataset}
		\label{fig:stepsa}
		\begin{tabular}{c c}
			\includegraphics[width=7cm]{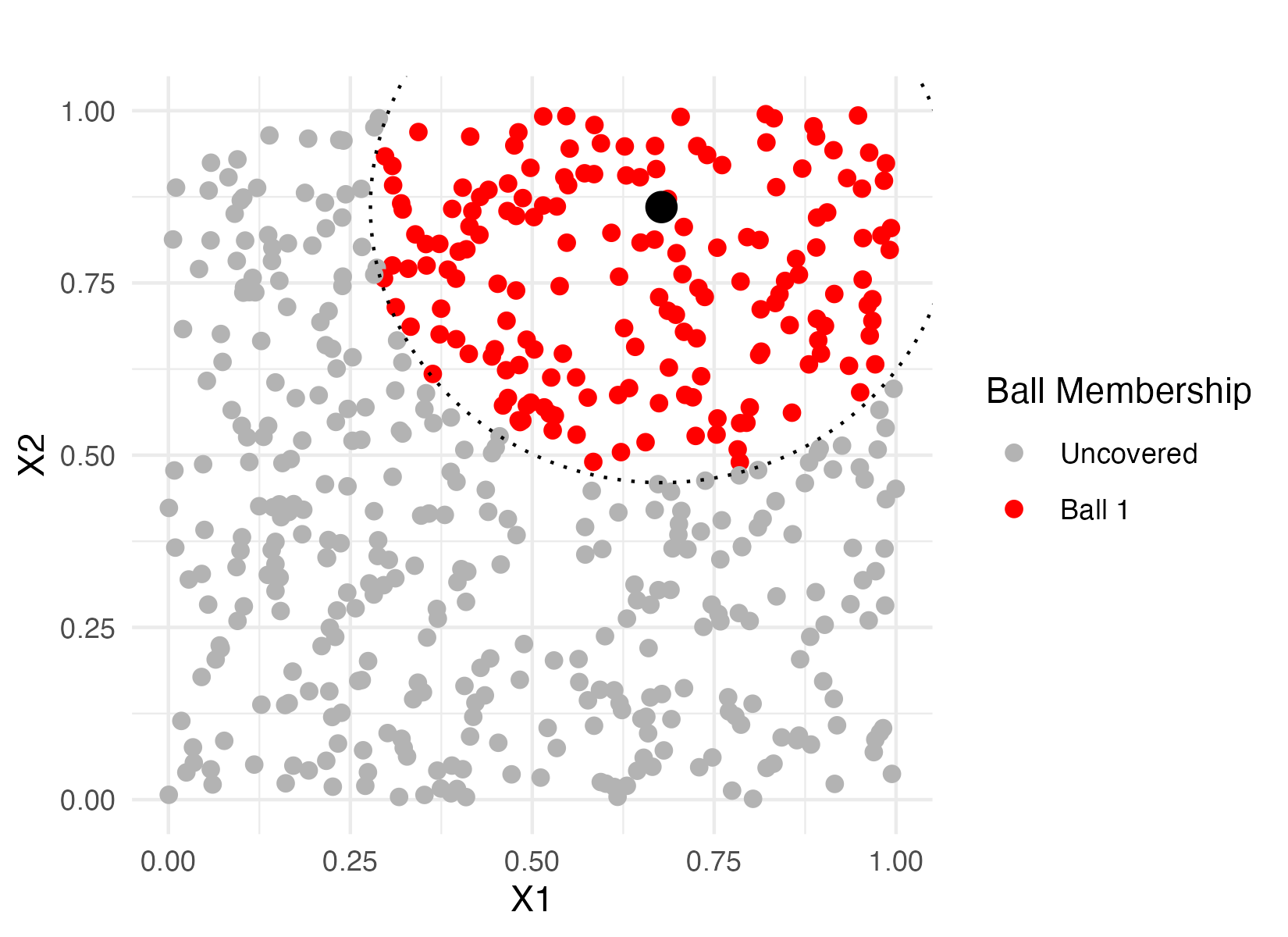}&
			\includegraphics[width=7cm]{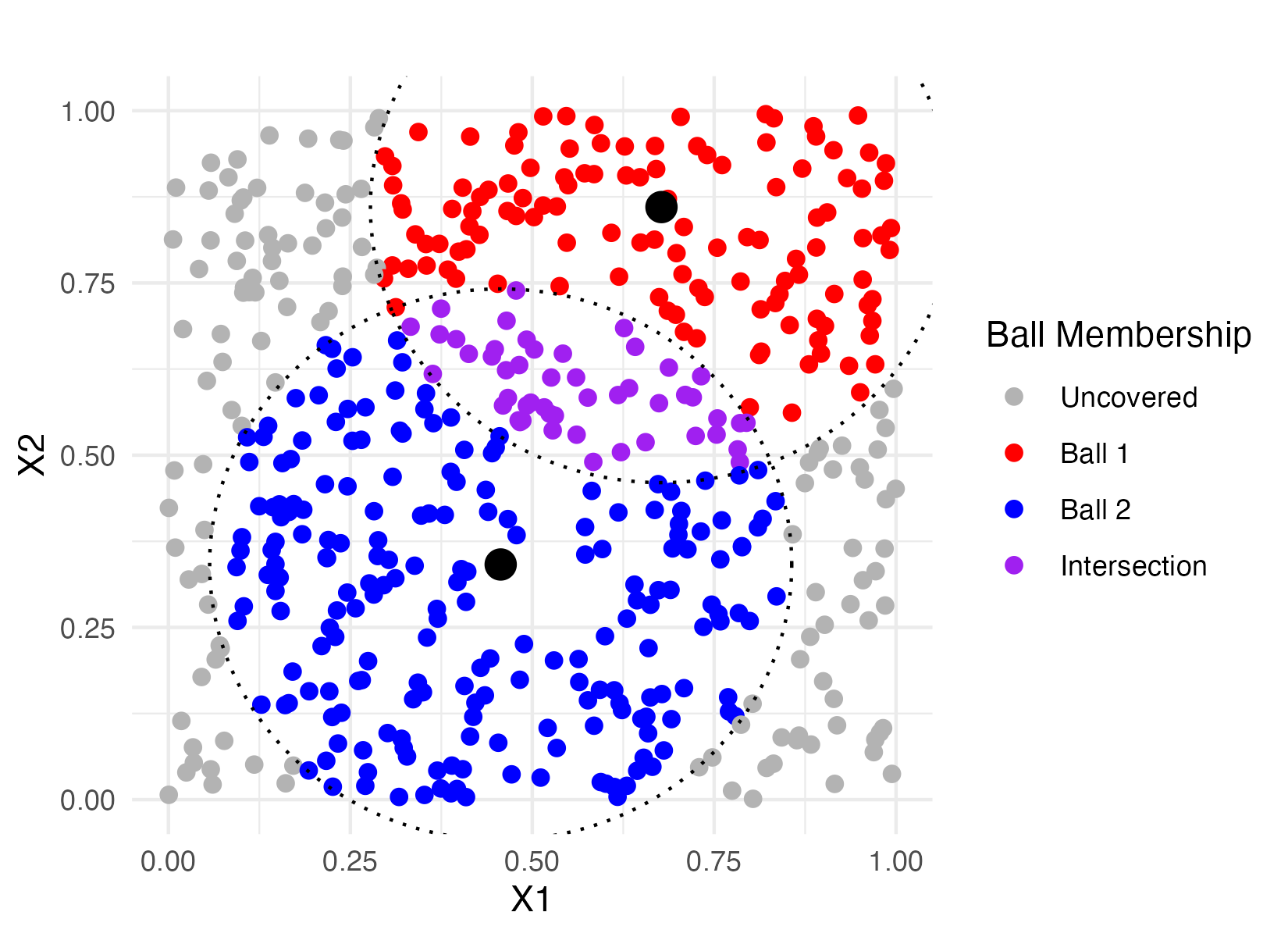}\\
			(a) First landmark, $l_1$ and Ball 1 & (b) Second landmark, $l_2$ and Ball 2 \\
			\includegraphics[width=7cm]{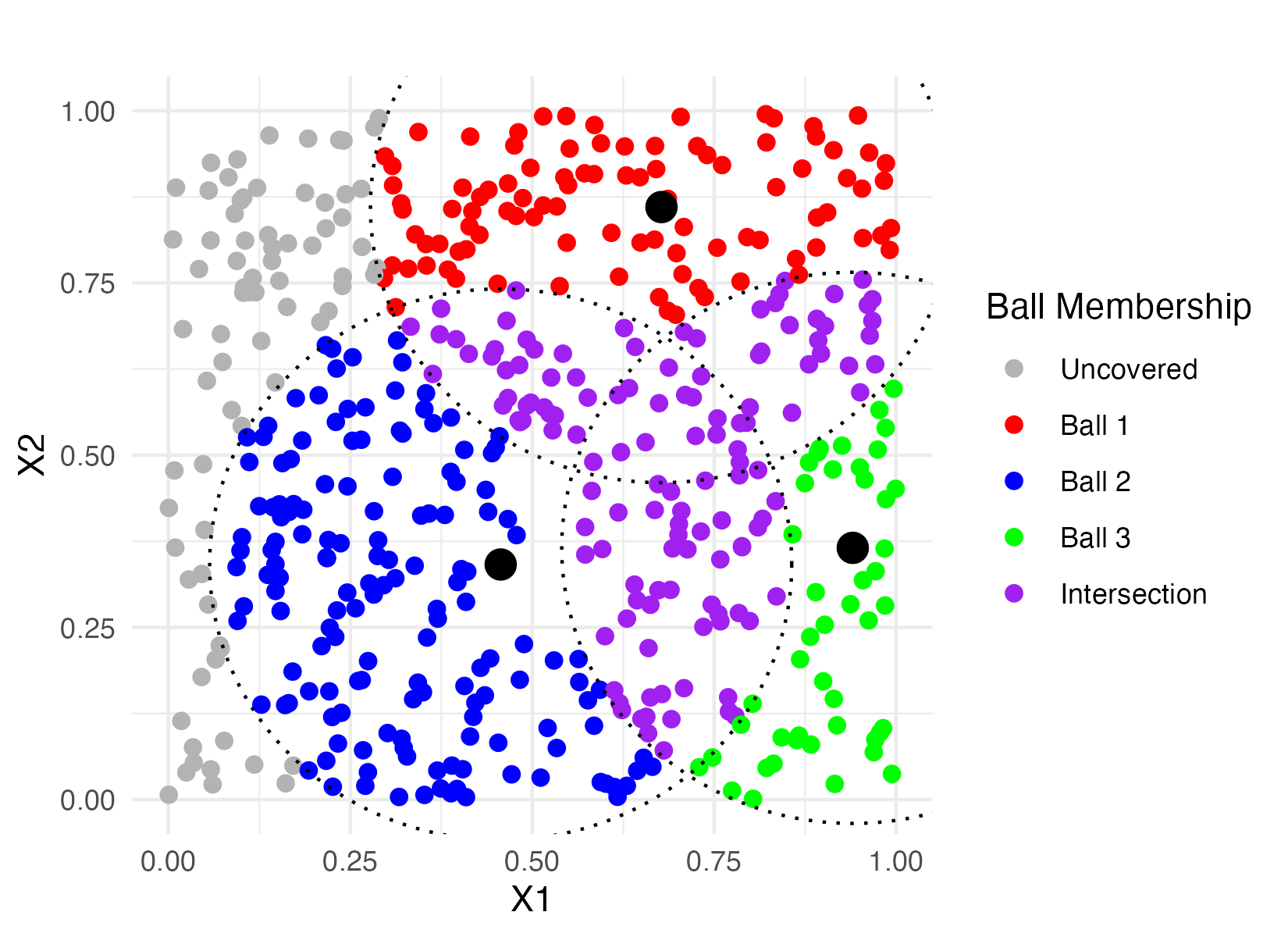}&
			\includegraphics[width=7cm]{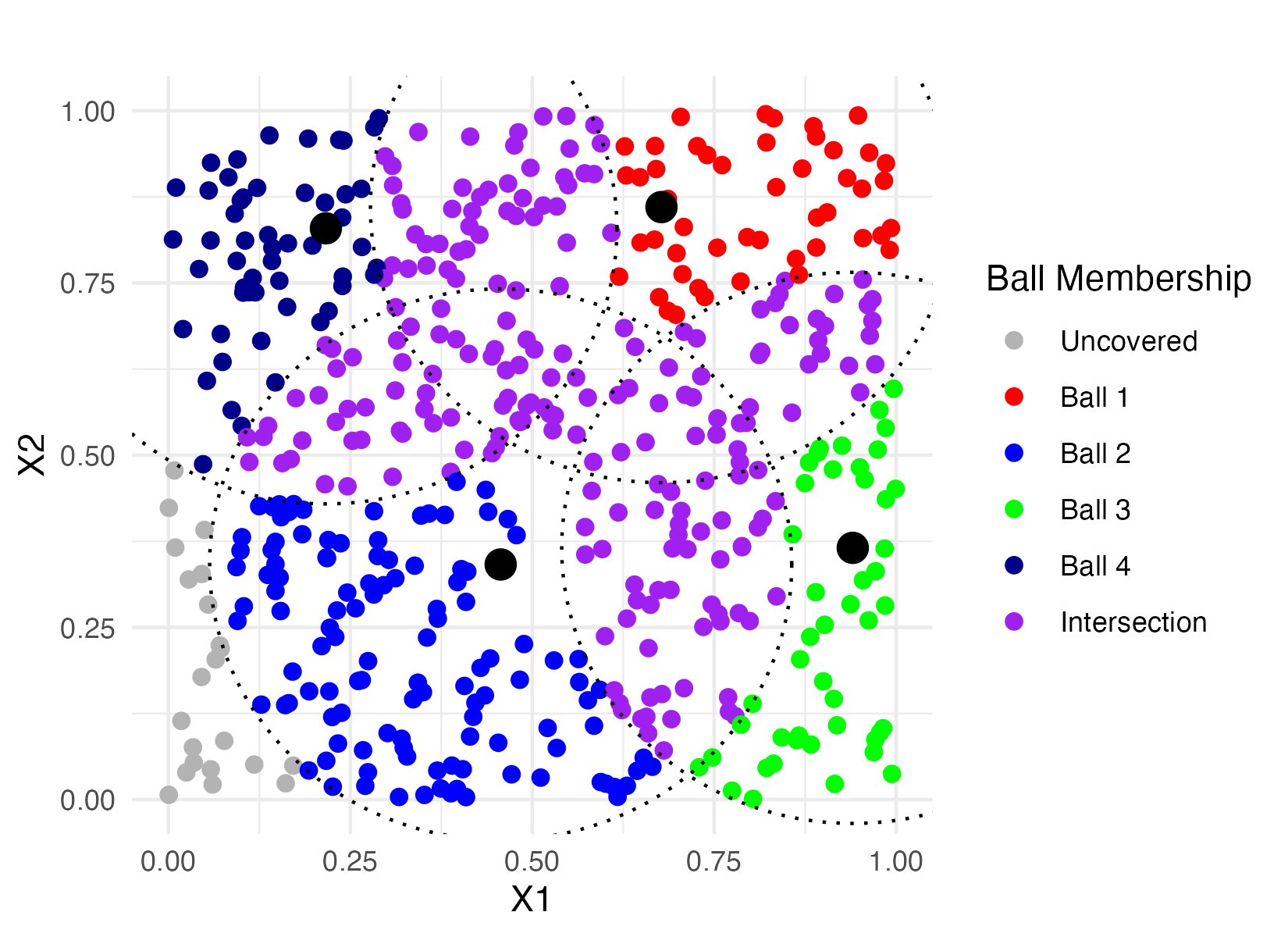}\\
			(c) Third landmark, $l_3$ and Ball 3 & (d) Fourth landmark, $l_4$ and Ball 4 \\
			\includegraphics[width=7cm]{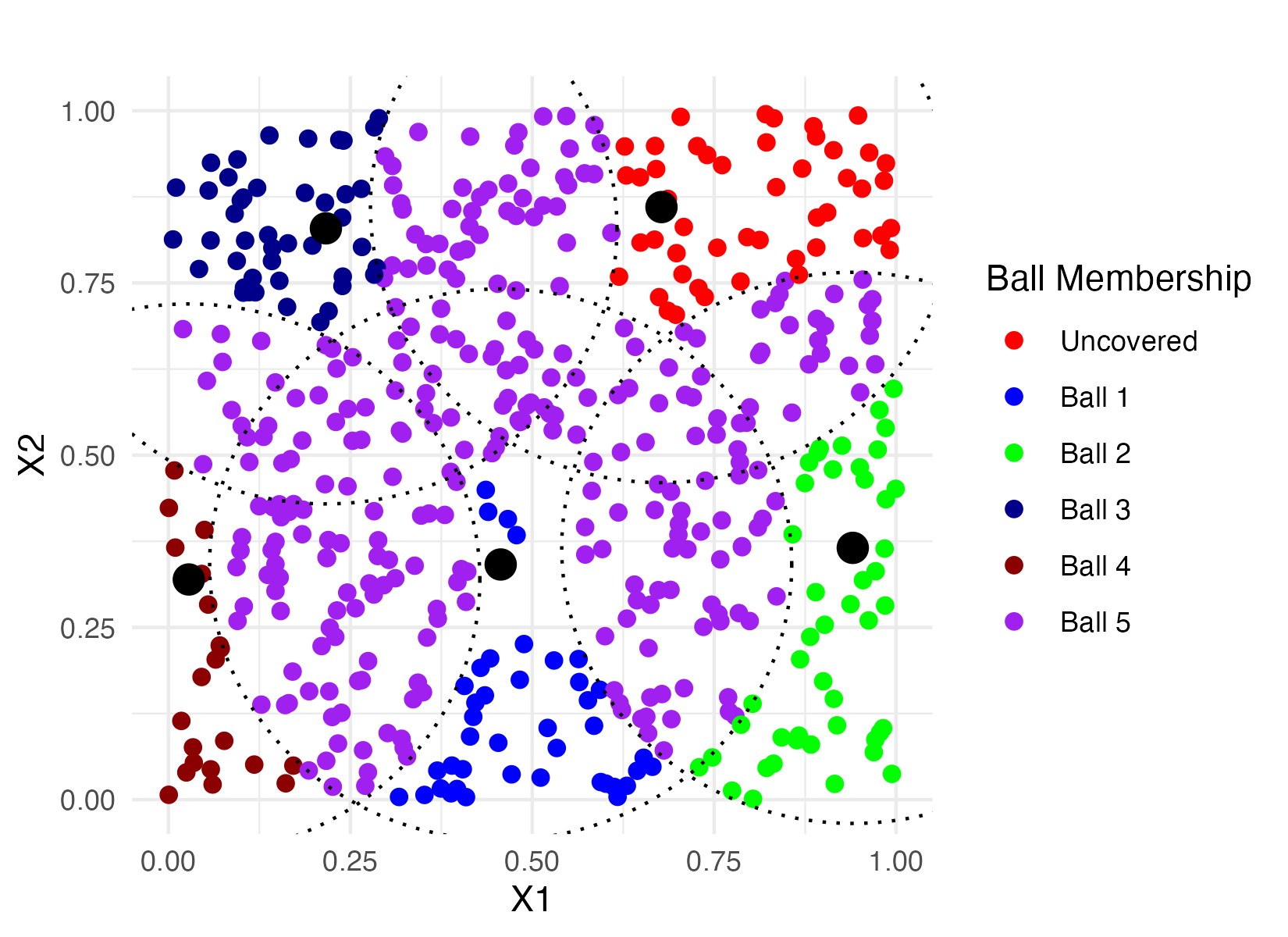}&
			\includegraphics[width=7cm]{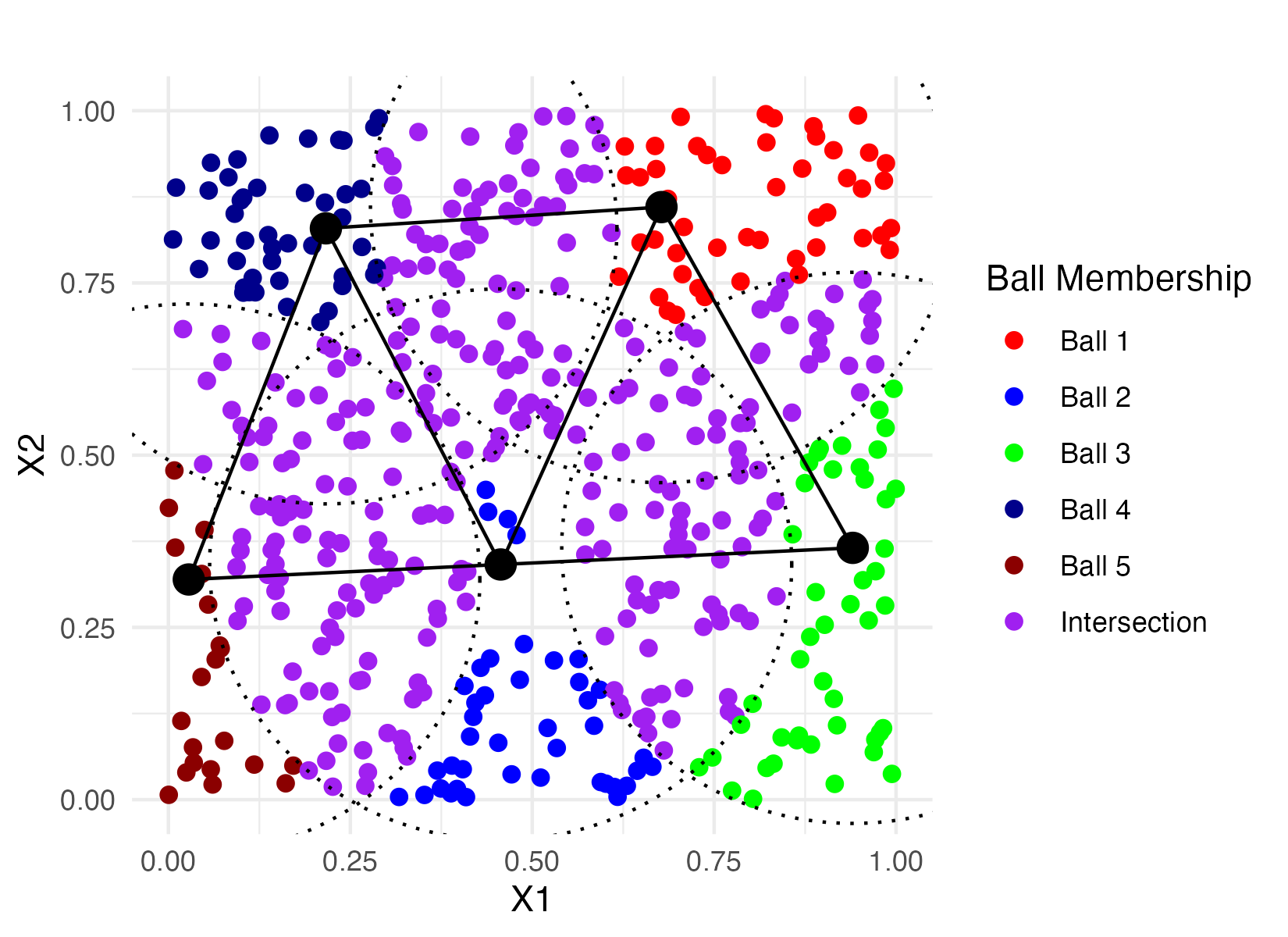}\\
			(e) Fifth landmark, $l_5$ and Ball 5 & (f) Edges for non-empty intersections \\
		\end{tabular}
	\end{center}
	\raggedright
	\footnotesize{Notes: Demonstration of the construction of a TDABM plot with random selection of landmark points. The dataset, $X$, comprises two variables, $X_1$ and $X_2$, which are drawn independently from $U[0,1]$. In each case a ball is drawn around a landmark point and all points within that ball are added to the covered set. Light grey points represent the uncovered set. After the selection of the 5th landmark, all points are covered. Edges are drawn between any pair of landmarks with a non-empty intersection.}
\end{figure}

Step 1 is to choose a point at random from the dataset. This point becomes the first landmark of the TDABM plot. The landmark is labelled $l_1$. A ball of radius $\epsilon$ is drawn around $l_1$. Label this ball $B_1 (X, \epsilon)$  All of the points within the ball are considered as being covered by ball 1. Panel (a) of Figure \ref{fig:stepsa} shows $l_1$ as a black point. Ball 1 is shown as a dashed line. The points covered by Ball 1 are shown in red.

Step 2 is to select a second landmark, $l_2$, from the set of uncovered points. In panel (a) of Figure \ref{fig:stepsa}, the uncovered points are colored grey. The second selected data point is shown as a black dot in panel (b) of Figure \ref{fig:stepsa}. Again a ball of radius $\epsilon$ is drawn around $l_2$. There is now a second ball $B_2(X,\epsilon)$ to add to the set of balls. The combined set of balls can now be defined as $B(X,\epsilon)$.  Panel (b) of Figure \ref{fig:stepsa} shows both balls and their intersection. Points in the intersection are in both balls 1 and 2. Uncovered points continue to be shown in light grey. As the coloration of the balls attests, R is aware which datapoints are within which ball.

Step 3 is the choice of a third landmark, $l_3$ from the uncovered set. Again a ball, $B_2(X,\epsilon)$, is created around $l_3$. There are now more points which are in the intersections of the balls. Panel (c) of Figure \ref{fig:stepsa} shows the construction with ball 3 present. The algorithm continues to select landmarks at random from the uncovered set until all points are covered. Figure \ref{fig:stepsa} shows the process with the selection of landmark $l_4$ in panel (d) and finally $l_5$ in panel (e). As each ball is added the set of light grey uncovered points shrinks.

The balls have numbers owing to the order in which they are created. Ball 1 is the ball surrounding the first landmark selected, etc. Because the ball numbers link to the random selection of points, they have no interpretation as numbers. Ball 1 has no superiority, or inferiority, to Ball 2 by fact of having a lower ball number. The numbers are provided to discuss the balls only. For example, if discussing the members of a single ball, it is known that those data points are similar in $X$. The ball number facilitates the discussion of the collection of similar points. Ball numbering gains further relevance when the balls are given a coloration, as is discussed later in this guide.

In order to represent the dataset further it is necessary to know the relative location of the balls. The final essential step of the algorithm is to draw edges between the landmarks of balls that have a non-empty intersection. Figure \ref{fig:stepsa} shows the network constructed by the edges within panel (b). We end the theoretical demonstration of the TDABM algorithm here.

\section{Ball Mapper as Implemented in R}
\label{sec:art}

There are small differences in the construction of the TDABM graph in R, compared to the theoretical construction. The differentials are explored in the first subsection here. The next step is to consider the inference that can be drawn from TDABM graphs. A subsection progressing from the basic network to the TDABM representation used in the literature is provided. 

\subsection{Landmark Selection and Graph Construction}

\begin{figure}
	\begin{center}
		\caption{Selection by First Uncovered Point}
		\label{fig:stepsb}
		\begin{tabular}{c c}
			\includegraphics[width=7cm]{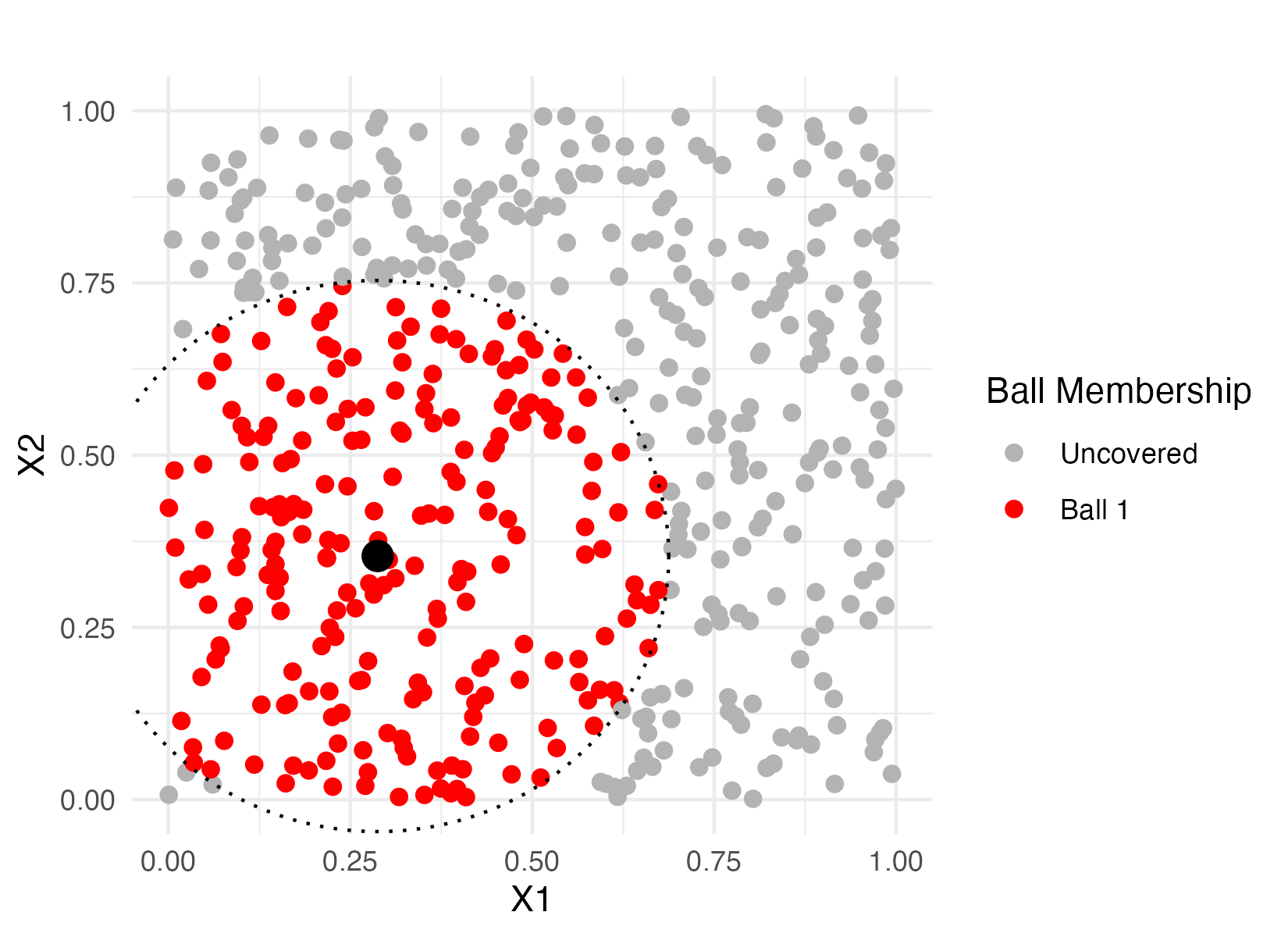}&
			\includegraphics[width=7cm]{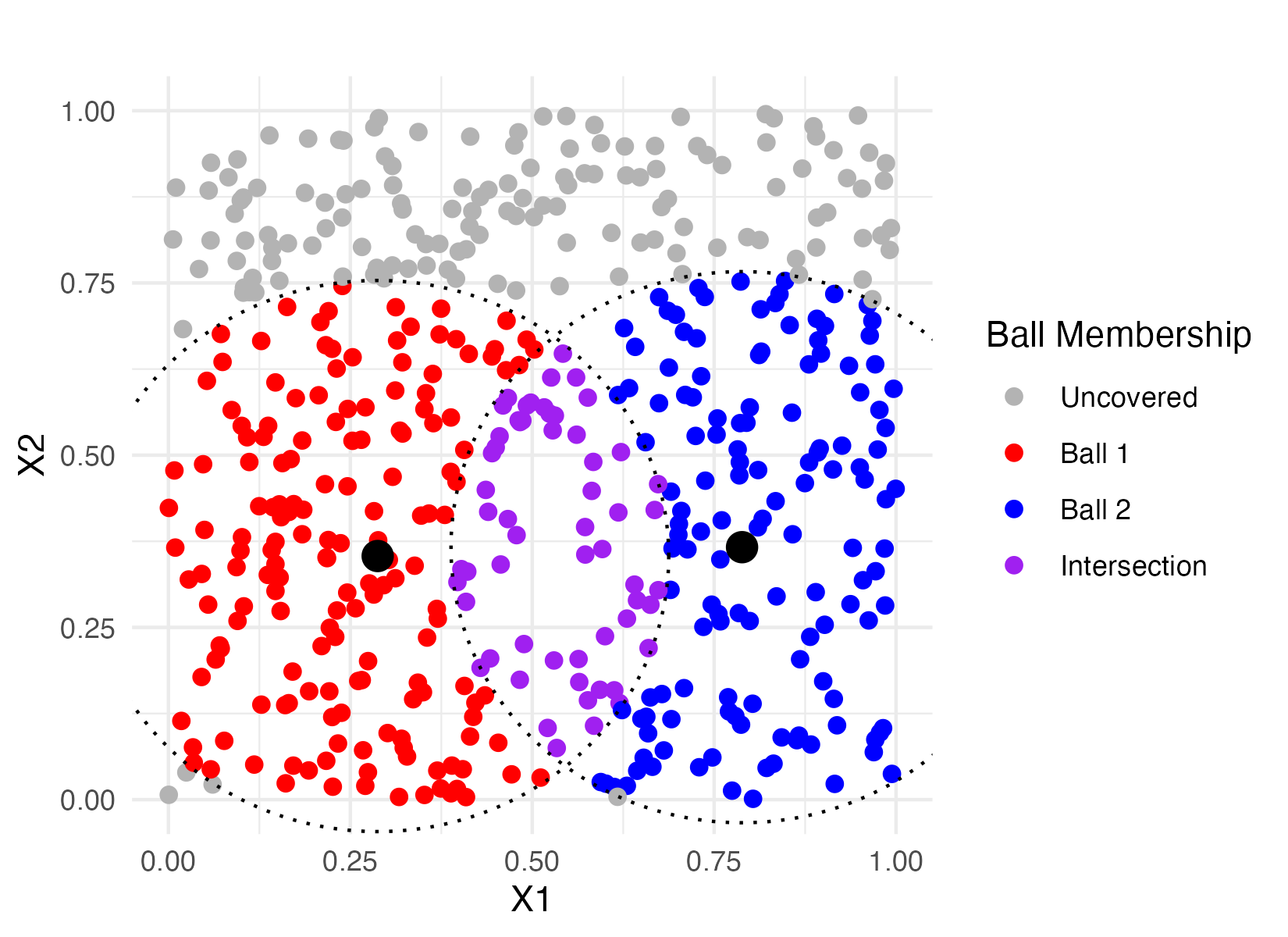}\\
			(a) First landmark, $l_1$ and Ball 1 & (b) Second landmark, $l_2$ and Ball 2 \\
			\includegraphics[width=7cm]{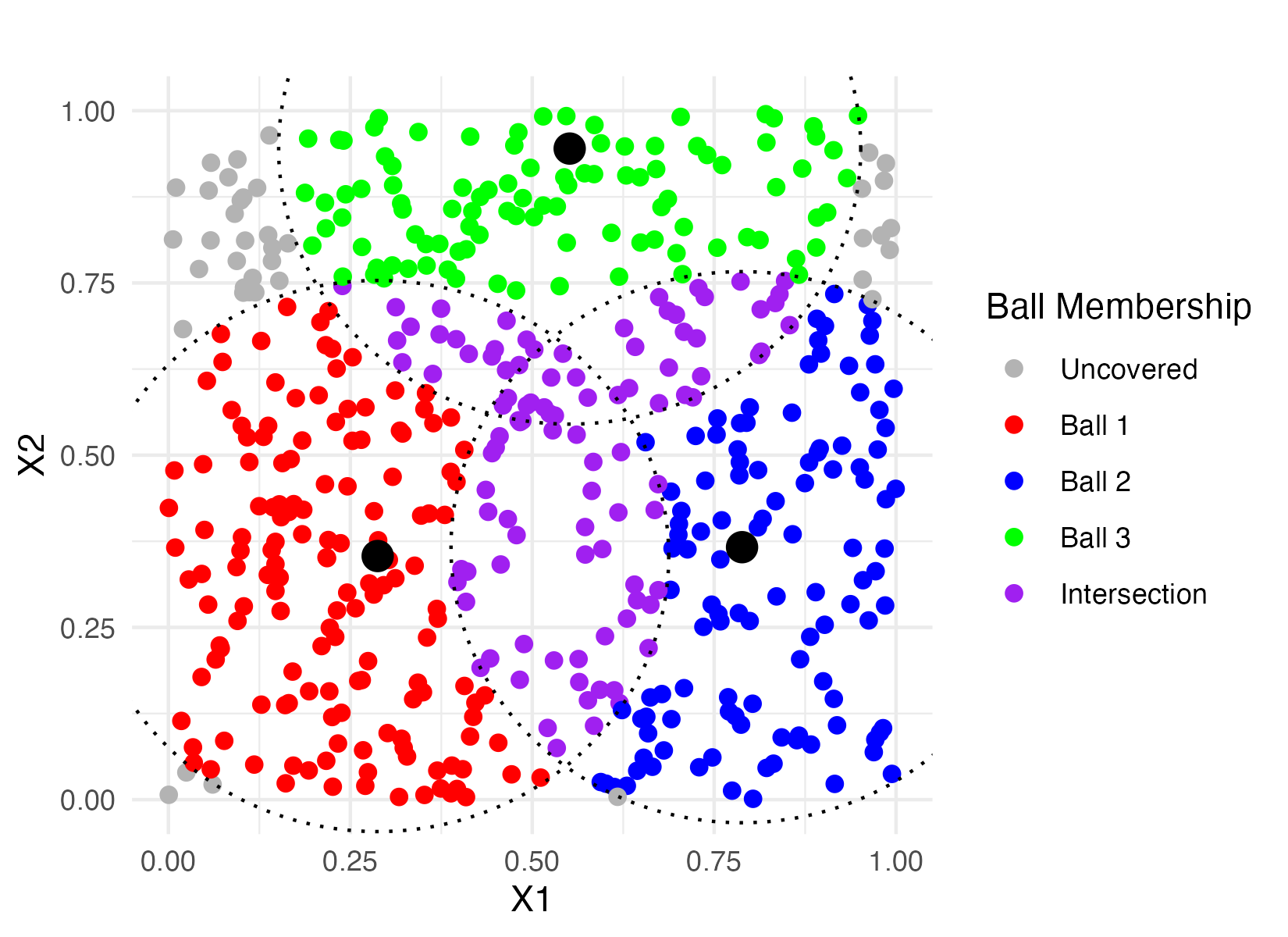}&
			\includegraphics[width=7cm]{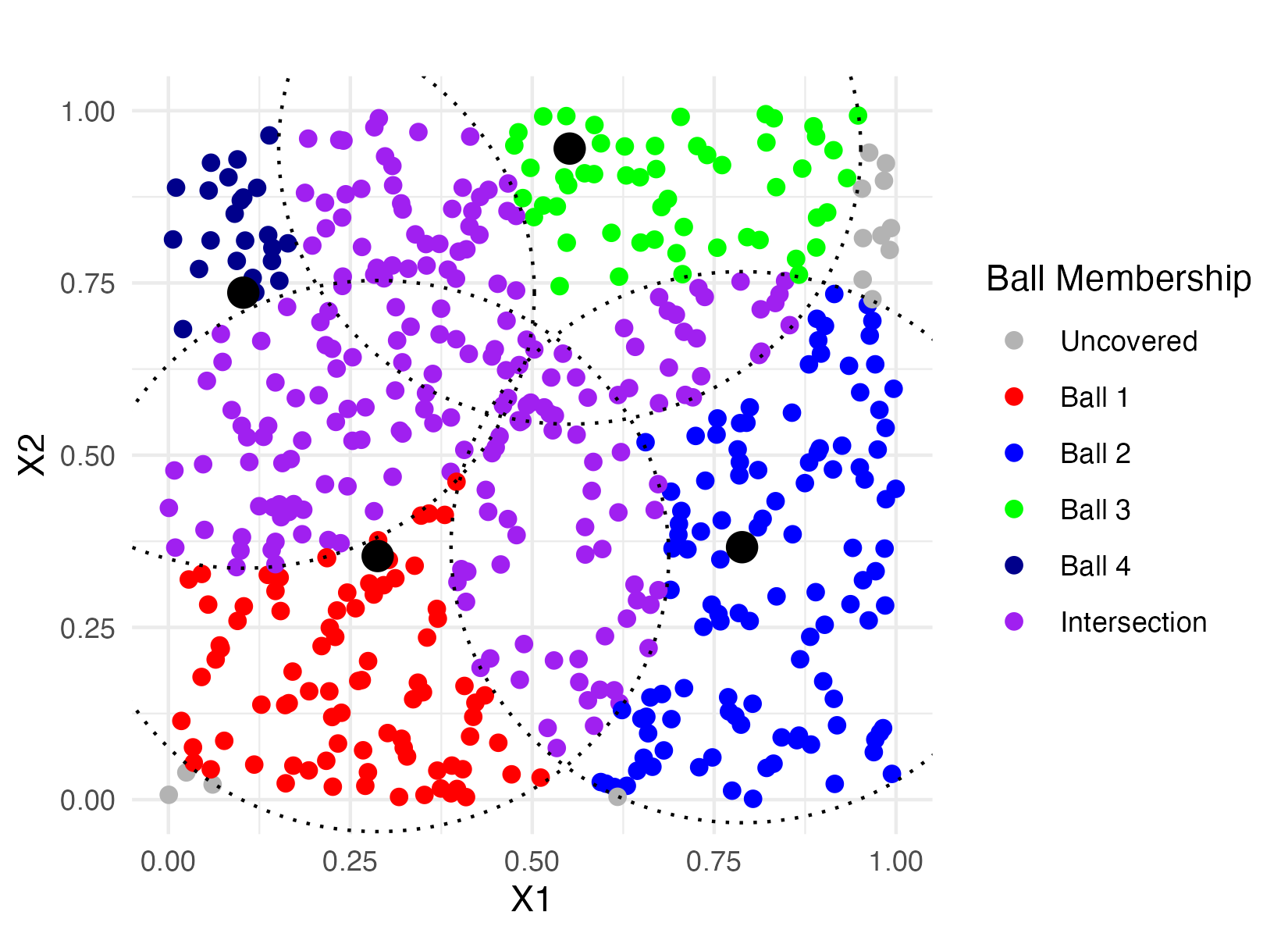}\\
			(c) Third landmark, $l_3$ and Ball 3 & (d) Fourth landmark, $l_4$ and Ball 4 \\
			\includegraphics[width=7cm]{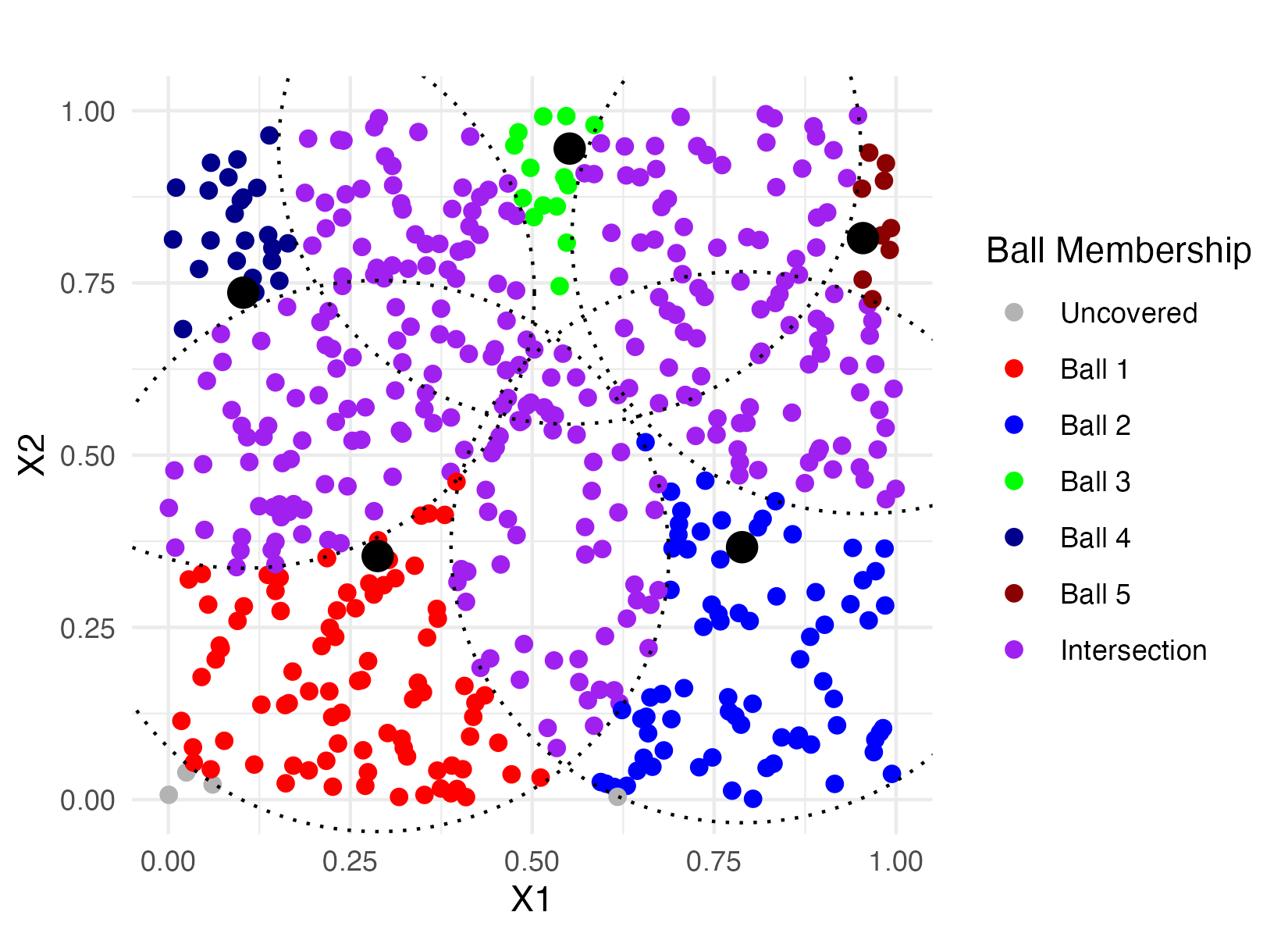}&
			\includegraphics[width=7cm]{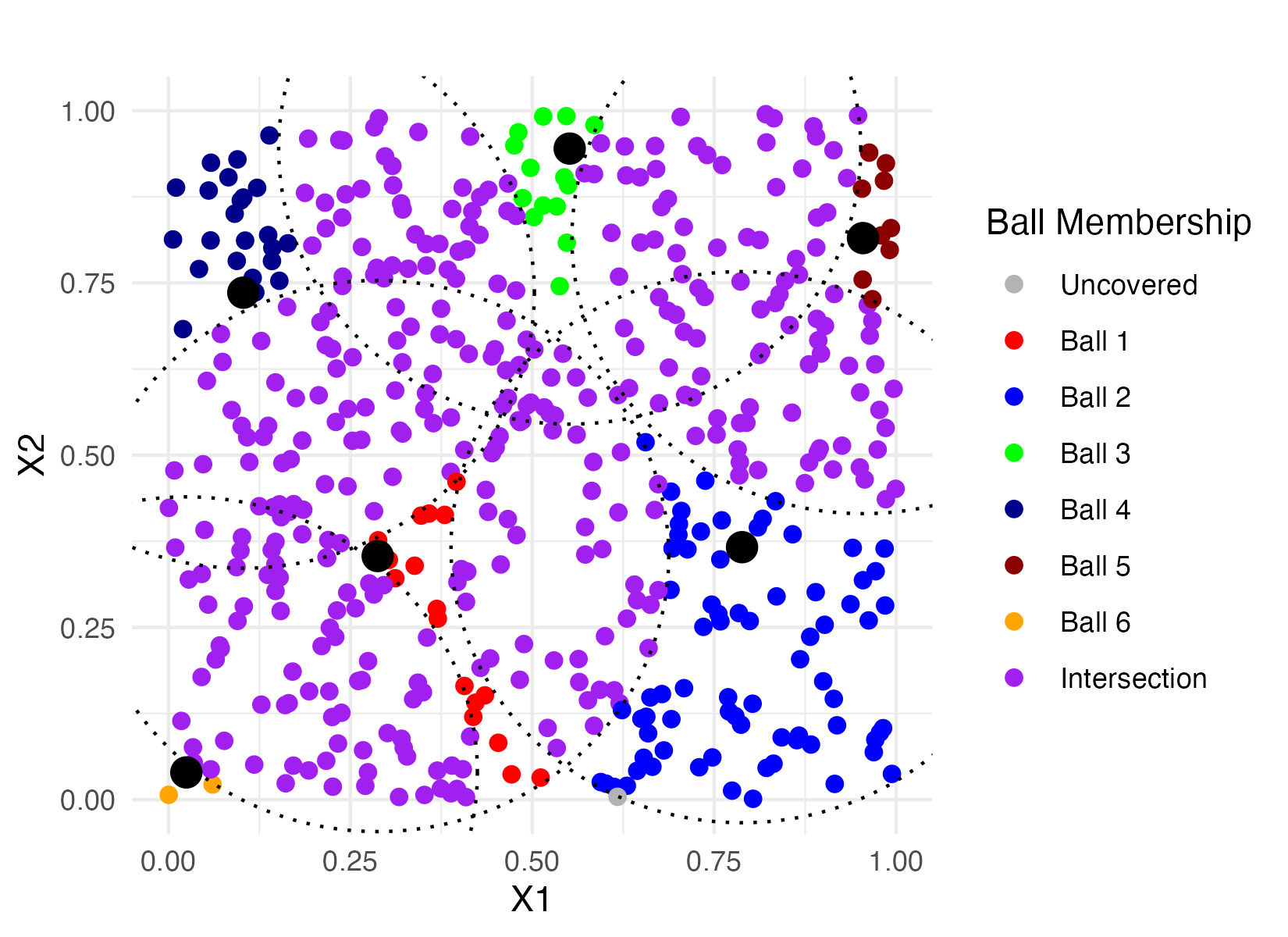}\\
			(e) Fifth landmark, $l_5$ and Ball 5  & (f) Sixth landmark, $l_6$ and Ball 6 \\
			\includegraphics[width=7cm]{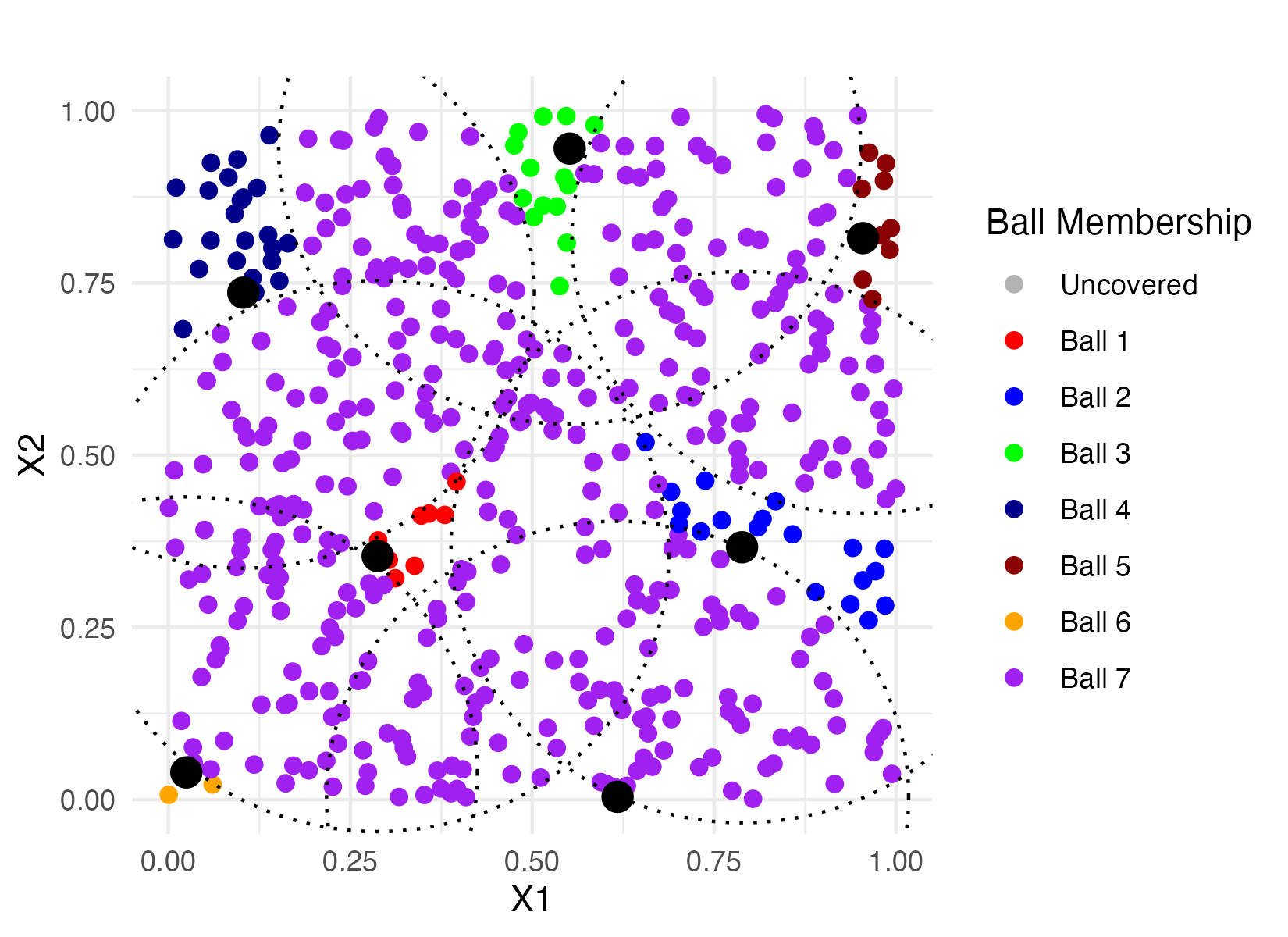}&
			\includegraphics[width=7cm]{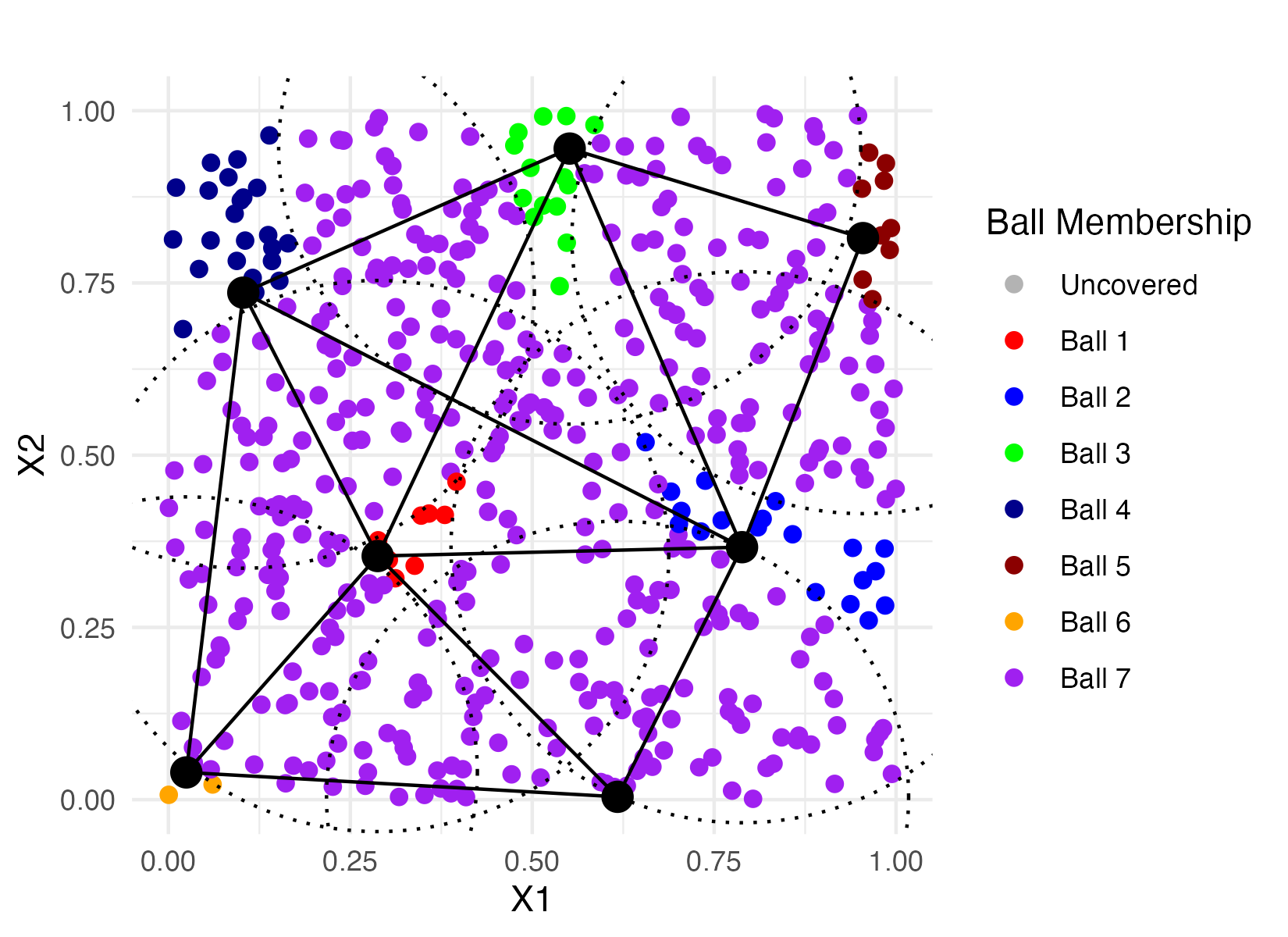}\\
			(g) Seventh landmark, $l_7$ and Ball 7 & (h) Edges connecting non-empty intersections  \\
		\end{tabular}
	\end{center}
	\raggedright
	\footnotesize{Notes: Demonstration of the construction as performed in the \texttt{BallMapper} package in R \citep{dlotko2019R}. Data is 500 points with two variables, $X_1$ and $X_2$. $X_1$ and $X_2$ are drawn independently from $U[0,1]$. Panels (a) to (g) show the iterative selection of the point from the uncovered set with the lowest ID in the uncovered set. In each case the points solely within one ball are colored to indicate their membership. Intersection points are colored purple. Panel (h) adds edges between the landmarks that have non empty intersections. Where relevant, light grey coloration represents the uncovered set. IDs are in ascending order for the dataset, $X$, which is supplied to the algorithm.}
\end{figure}

The R implementation of the TDABM algorithm is performed using the package \texttt{BallMapper} by \cite{dlotko2019R}. To allow the user to randomize the selection of landmarks, the R algorithm simply chooses the data point with the lowest index that is still within the uncovered set. Hence with a given seed you will always get the same landmarks. Likewise, by shuffling the order of the input data it is possible to obtain a different TDABM representation of the data. In this section I repeat the process from Figure \ref{fig:stepsa}, but by using first uncovered point as the landmark. Figure \ref{fig:stepsb} provides the results.

Figure \ref{fig:stepsb} shows the selection of the first landmark point, $l_1$ in panel (a). The points contained within $B_1(X,\epsilon)$ are shown in red. Panel (b) has the addition of the second landmark. We see the points which are only in $B_2(X,\epsilon)$ colored blue. The points in the intersection of the two balls are colored purple. This is the same as in Figure \ref{fig:stepsa}, but the landmarks selected are different. Continuing, the algorithm selects $l_3$, $l_4$, $l_5$, $l_6$ and $l_7$. Panels (c) to (g) show how the balls cover the full dataset. Panel (h) has the edges between any pair of landmarks that have a non-empty intersection. In this case the number of balls needed to cover the full dataset is 7. The 7 compares to just 5 balls in the previous case. \cite{dlotko2022topological}, \cite{rudkin2023economic}, \cite{rudkin2024return} and \cite{rudkin2024topology} all show that there are small variations in the number of balls obtained when repeating the algorithm multiple times. However, in all cases the inference drawn is consistent. 

Within the code for each step, the first task is to identify the uncovered points. Secondly, the distances between all data points and the selected landmark point are calculated. The points within the radius become the points in the ball that is being constructed. An example for constructing Ball 6 is seen in Box \ref{box:ball6}. A dataset of landmarks, here called $sel$ is appended with the new landmark, $l_6$. The next block of code is designed to isolate all of the intersections to create the purple coloration on the illustration. Finally, the \texttt{member} dummy is created for any point which is either in one of the balls, or in the intersection of more than one ball. There may be a more efficient way to obtain a simulation of the TDABM algorithm. However, the aim here is not to provide efficient code as I am instead to explain what the R function \texttt{BallMapper} from \cite{dlotko2019R} is doing under the hood.

\begin{mybox}[label=box:ball6]{R Code for Imitating Creation of Ball 6}
	When selecting a landmark, R will look for the lowest id in the uncovered set:
	\begin{lstlisting}[language=R]
		unc<-subset(df1,df1$member==0)
		minid<-min(unc$pt)
	\end{lstlisting}
	The landmark can be identified in the dataset and distances to that point measured:
	\begin{lstlisting}[language=R]
		selected_point6 <- df1[minid, c("X1", "X2")]
		df1$distance <- sqrt((df1$X1 - selected_point6$X1)^2 + (df1$X2 - selected_point6$X2)^2)
		df1$within_radius6 <- ifelse(df1$distance <= 0.4, 1, 0)
	\end{lstlisting}
    Convert the selected landmark point into a dataframe
	\begin{lstlisting}
		selected_point6<-as.data.frame(cbind(selected_point6$X1,selected_point6$X2))
		names(selected_point6)<-c("X1","X2")
		sel<-as.data.frame(rbind.data.frame(sel,selected_point6))
	\end{lstlisting}	
	Create the intersections and membership
	\begin{lstlisting}
		df1$ball6<-ifelse(df1$within_radius6==1&df1$ball1==0&df1$ball2==0& df1$ball3==0&df1$ball4==0&df1$ball5==0&df1$intersection99==0,6,0)
		df1$intersection5<-ifelse(df1$within_radius6==1&df1$within_radius==1,99,0)
		df1$intersection5<-ifelse(df1$within_radius6==1&df1$within_radius2==1,99, df1$intersection5)
		df1$intersection5<-ifelse(df1$within_radius6==1&df1$within_radius3==1,99, df1$intersection5)
		df1$intersection5<-ifelse(df1$within_radius6==1&df1$within_radius4==1,99, df1$intersection5)
		df1$intersection5<-ifelse(df1$within_radius6==1&df1$within_radius5==1,99, df1$intersection5)
		df1$intersection99<-df1$intersection+df1$intersection2+df1$intersection3+ df1$intersection4+df1$intersection5
		df1$intersection99<-ifelse(df1$intersection99>0,99,0)		
		df1$member<-df1$ball1+df1$ball2+df1$ball3+df1$ball4+df1$ball5+df1$ball6+ df1$intersection99
		df1$member<-ifelse(df1$member<8,df1$member,100)
	\end{lstlisting}
	This example has produced the 6th landmark within the example. There may be more efficient coding mechanisms to obtain the same result.
\end{mybox}

\subsection{Interpreting TDABM Graphs}

\begin{figure}
	\begin{center}
		\caption{Annotating TDABM Plots}
		\label{fig:stepsc}
		\begin{tabular}{c c}
			\includegraphics[width=7cm]{demo201.png}&
			\includegraphics[width=7cm]{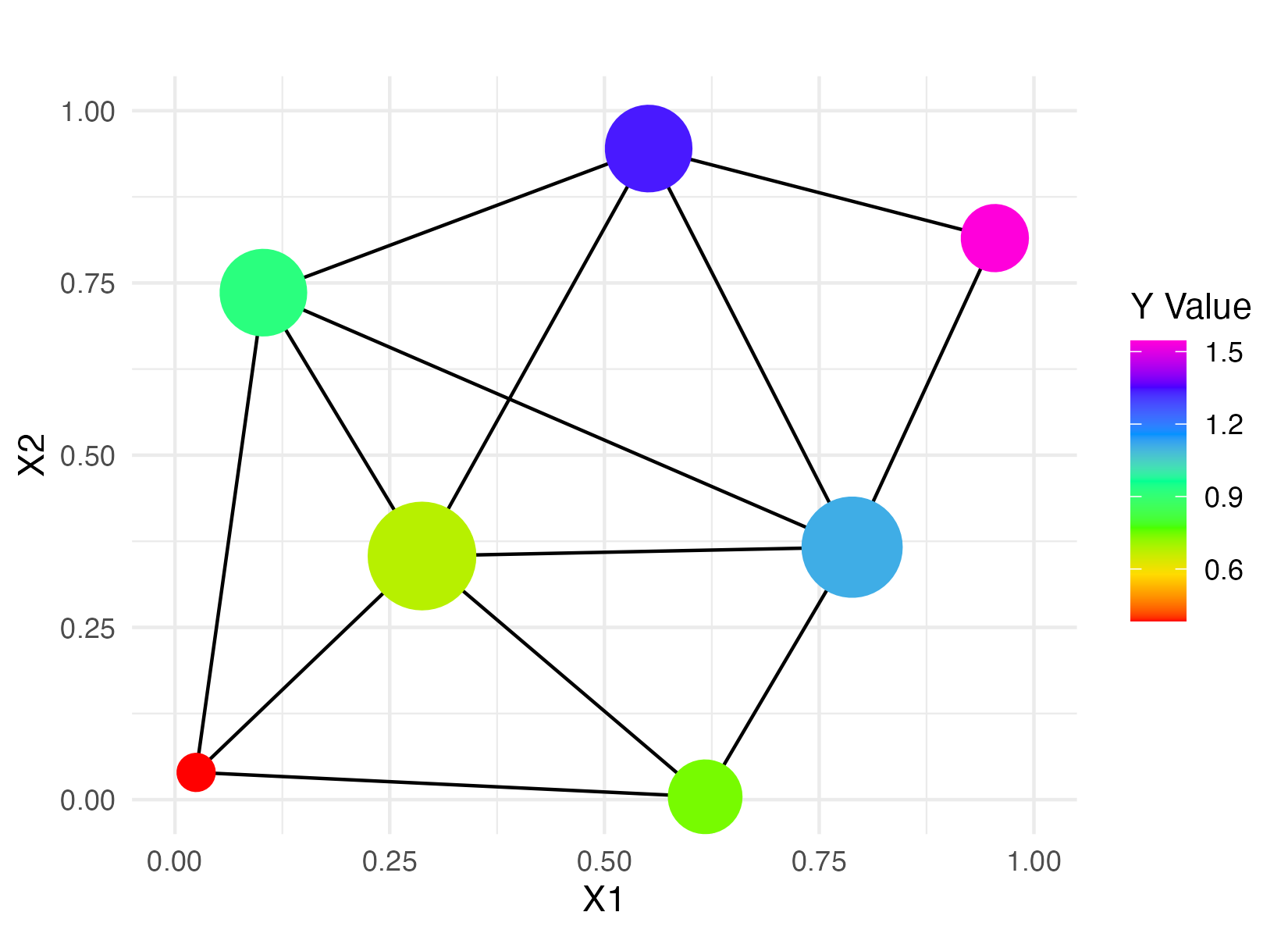}\\
			(a) Raw dataset colored by $Y$ & (b) \texttt{BallMapper} style representation \\
			\includegraphics[width=7cm]{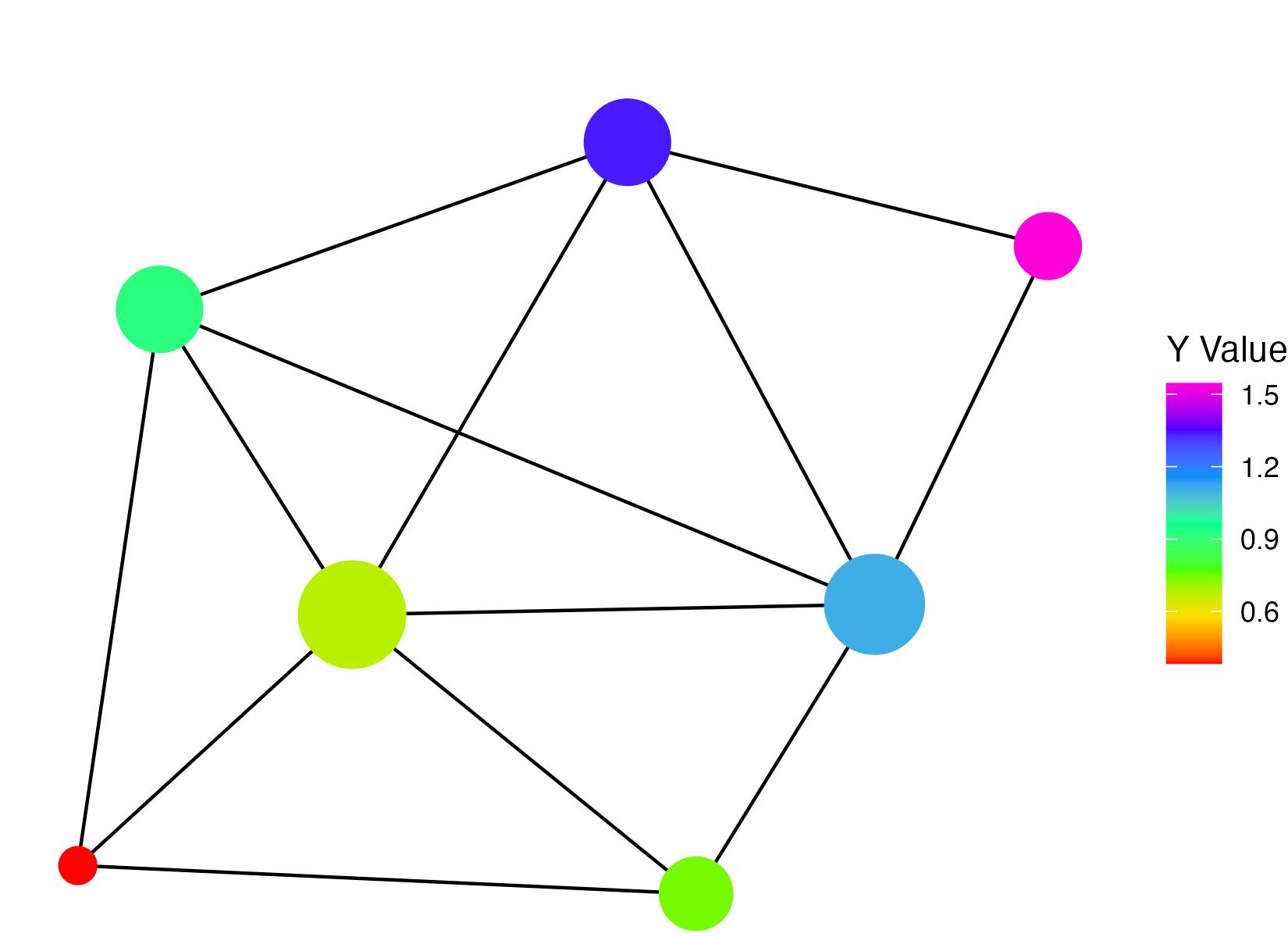}&
			\includegraphics[width=7cm]{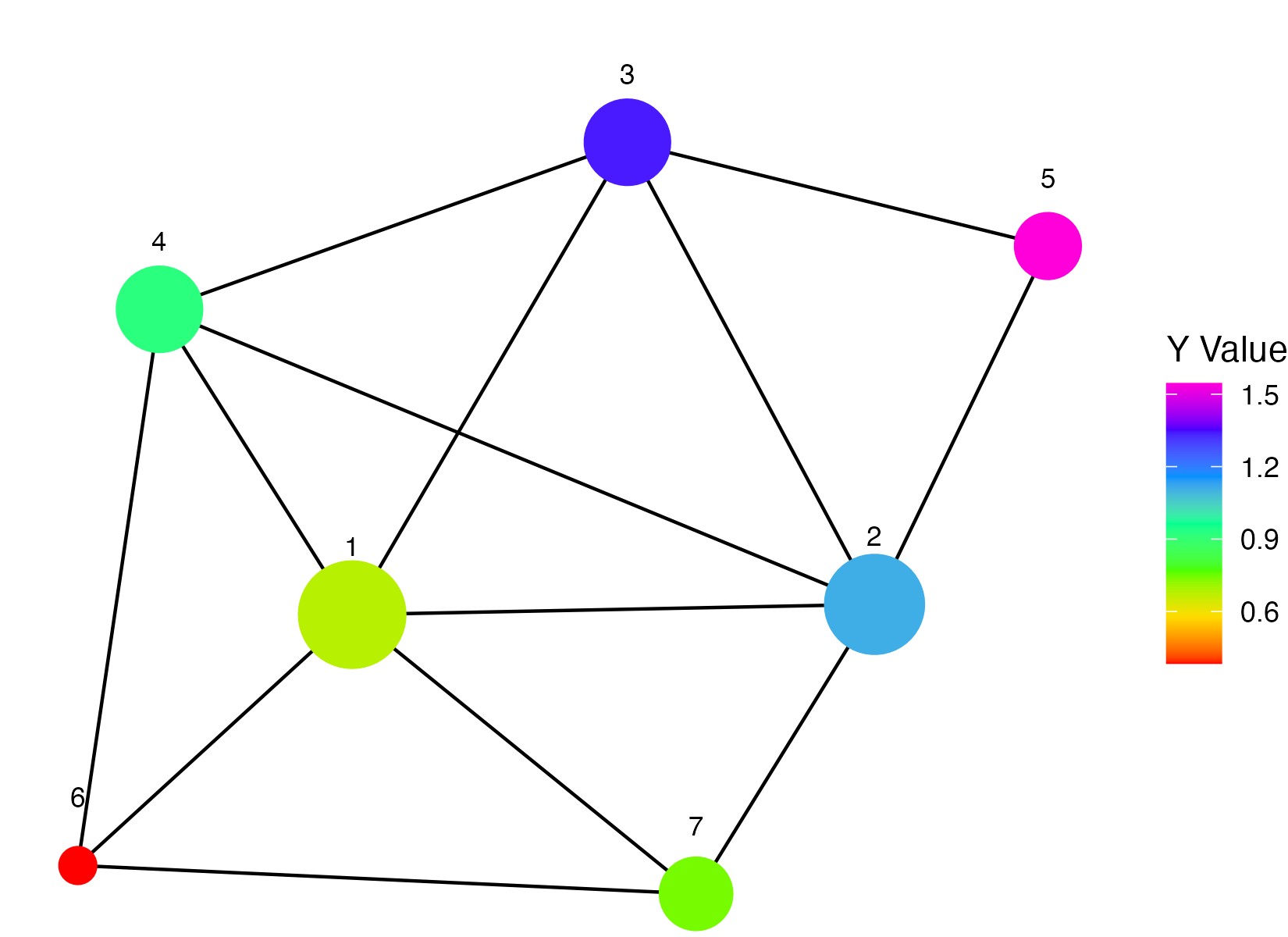}\\
			(c) Removal of Axes & (d) Labelling of Balls \\
		\end{tabular}
	\end{center}
	\raggedright
	\footnotesize{Notes: Demonstration of the construction of a TDABM plot for the dataset shown in panel (a). Panel (a) shows the $N=500$ data points colored according to the value $y_i = x_{1i}+x_{2i}$. Variables $X_1$ and $X_2$ are drawn independently from $U[0,1]$. Panel (b) shows the balls as constructed in Figure \ref{fig:stepsb}. The ball size corresponds to the number of points within the ball normalised onto an interval of sizes between 5 and 15. Coloration of the balls is according to the average value of $y_i$ across all points within the ball. Panel (c) shows the same plot without the axes. Panel (d) has numbers corresponding to the order in which the landmarks are selected.}
\end{figure}

To this point, the demonstration has focused on the construction of the TDABM graph. Because TDABM knows which data points are in which ball, it is possible to enhance the visualization to represent the density of the joint distribution of $X$. It is also possible to map $Y$ across the space using the TDABM visualization. Figure \ref{fig:stepsc} demonstrates the process of getting from the raw data to the final representation.

Figure \ref{fig:stepsc} panel (a) repeats the dataset picture from Figure \ref{fig:raw} as a comparator for the other panels. Panel (b) is based upon panel (h) of Figure \ref{fig:stepsb}. Here the landmarks have been resized to reflect the number of data points that appear within each ball. The sizes are normalized to ensure that the representation of the ball does not become too large. Instead there is a minimum size associated with the smallest number of points and a maximum associated with the largest number of points.  By resizing the data, an impression of the density of the point cloud in the area surrounding each landmark is provided.

The coloration is based upon the average value of $Y$ across all of the points within the ball. Here, advantage is taken of the fact that the algorithm knows exactly which points are within each ball. In the code dummies have been created for points within the fixed radius of each landmark. To obtain the average value of $Y$, it is simply a case of subsetting to just those points in the ball and finding the mean. Code is omitted from this section for brevity. Because TDABM plots are abstract, coloration by the $X$ axes can be used to understand how each varies across the TDABM plot. In this case, the balls remain at the correct co-ordinates in $X_1, X_2$ space such that coloration by either $X_1$ or $X_2$ is not necessary.

Once the coloration has been added the ball numbers become useful. In this example, the ball numbers are added manually in panel (d) of Figure \ref{fig:stepsc}. Now it is possible to say that Ball 1 has the highest value of the coloration function. Ball 5 has the lowest number of observations contained within, Ball 2 is connected to Ball 3 etc. It is through the discussion of ball numbers that more is understood about the structure of the data. When the coloration of the TDABM is generated from real-world data the discussion of the colors of balls has more meaning. Low, or high, coloration balls amonst otherwise high, low, balls would indicate an important variation in the outcome surface that warrants attention. The ball numbers have no further interpretation beyond being devices to describe the plot.

\subsection{TDABM Graph}

The previous subsections have documented the process through which the TDABM graph is constructed. However, when implementing TDABM in R, a single command is used. Code is provided in Box \ref{box:dprep} for the production of the datasets from the \texttt{df1} dataset that has been built in previous subsections. Secondly, Box \ref{box:tdabm} shows the implementation of the TDABM algorithm.

\begin{mybox}[label=box:dprep]{Preparing Data for TDABM}
	First produce the requisite data sets for the axes and outcomes:
	\begin{lstlisting}[language=R]
		xd<-as.data.frame(cbind.data.frame(df1$X1,df1$X2))
		yd<-as.data.frame(df1$Y)
	\end{lstlisting}
	We bind the two variables together to produce a \texttt{data.frame} object:
	\begin{lstlisting}[language=R]
		names(xd)<-c("X1","X2")
		names(yd)<-"Y"
	\end{lstlisting}
\end{mybox}

Having created a dataframe for the axes, a dataframe for the outcome and chosen a radius, it is possible to generate the TDABM graph. The code for generating the TDABM graph is provided in Box \ref{box:tdabm}. The simplicity of the \texttt{BallMapper} function itself is clear.

\begin{mybox}[label=box:tdabm]{R Code for Implementing TDABM}
	Having constructed the axes as \texttt{xd} and outcome as \texttt{yd}, the TDABM algorithm with radius $\epsilon=0.4$ is implemented using
	\begin{lstlisting}[language=R]
		bm1<-BallMapper(xd,yd,0.4)
	\end{lstlisting}
	The result is a BallMapper object, \texttt{bm1}.
\end{mybox}

\begin{figure}
	\begin{center}
		\caption{TDABM Plot}
		\label{fig:tdabm}
		\includegraphics[width=12cm]{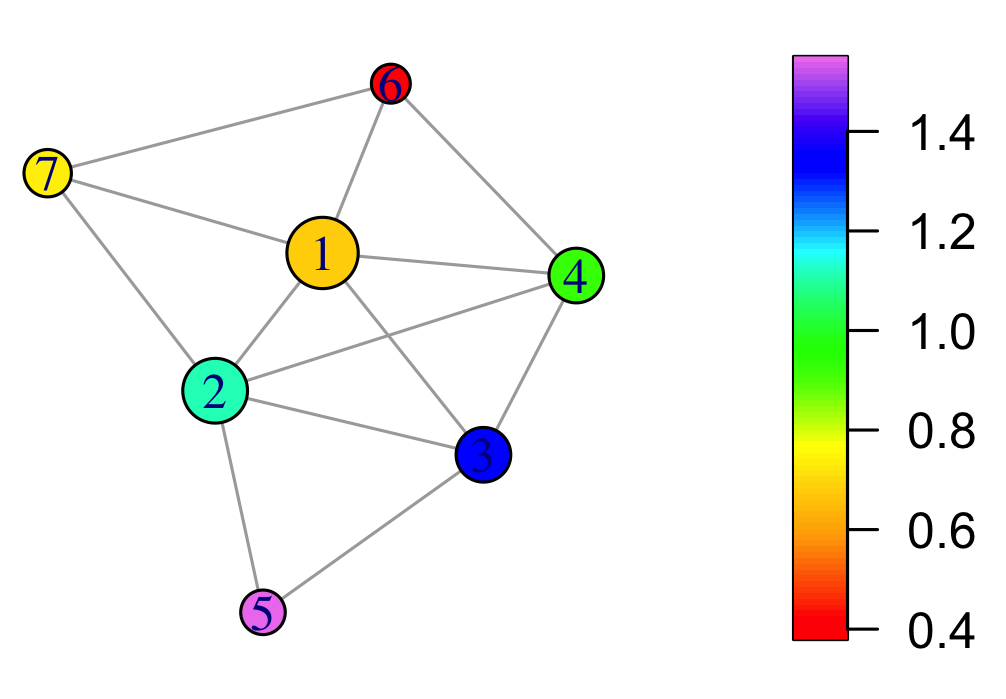}
	\end{center}
		\raggedright
	\footnotesize{Notes: TDABM plot of a bivariate dataset $X$ in which $X_1$ and $X_2$ are drawn independently from $U(0,1)$. TDABM graph constructed with ball radius 0.4. The generation process for the outcome $Y$ has $y_i = x_{i1} + x_{i2}$. Coloration is according to the average value of $Y$ within each ball. TDABM implemented using the R package \texttt{BallMapper} by \cite{dlotko2019R}.}
\end{figure}

Plotting the TDABM object is done using the \texttt{ColorIgraphPlot} command. The plotting makes use of the \texttt{igraph} package of \cite{igraphp}, based on the paper \cite{csardi2006igraph}. The \texttt{igraph} package is also the workhorse package for social network analysis in R. For the purposes of plotting TDABM, the ability to recreate the TDABM graph as a network is the key advantage. When plotting more than 2 dimensions, the result is necessarily abstract. The parameter \texttt{seed\_for\_plotting} allows the user to change the representation for readability without altering the structure. Box \ref{box:igraph} shows the code for plotting the \texttt{bm1} object constructed following Box \ref{box:tdabm}. The resulting TDABM graph is shown in Figure \ref{fig:tdabm}.

\begin{mybox}[label=box:igraph]{R Code for Generating a Simple TDABM Plot}
	The TDABM object, \texttt{bm1}, can be plotted in R using:
	\begin{lstlisting}[language=R]
		ColorIgraphPlot(bm1,seed_for_plotting=123)
	\end{lstlisting}
\end{mybox}

Many similarities can be seen between the plot of Figure \ref{fig:tdabm} and panel (d) of Figure \ref{fig:stepsc}. The largest balls are numbers 1 and 2, as was seen in Figure \ref{fig:stepsc}. Ball 6 has the lowest value of the coloration function, whilst Ball 5 has the highest coloration. In the plot of panel (a), it can be seen that Ball 5 is to the North East where $X_1 + X_2$ is at the maximum. Ball 6 is located to the lower left where $X_1 + X_2$ is at its lowest. Because the artificial simulation of TDABM did not make the plot abstract, the mapping from the balls to the original dataset is easier to understand. However, the abstract representation is still easy to follow. 

\begin{figure}
	\begin{center}
		\caption{TDABM Plot Axes}
		\label{fig:tdabm2}
		\begin{tabular}{c c}
			\includegraphics[width=7cm]{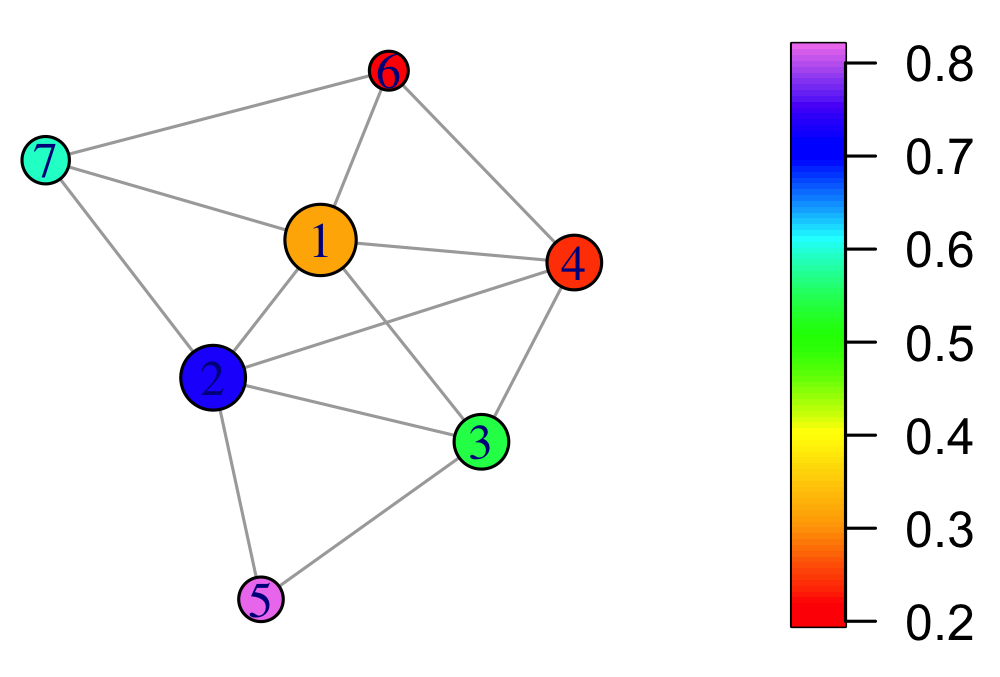}&
			\includegraphics[width=7cm]{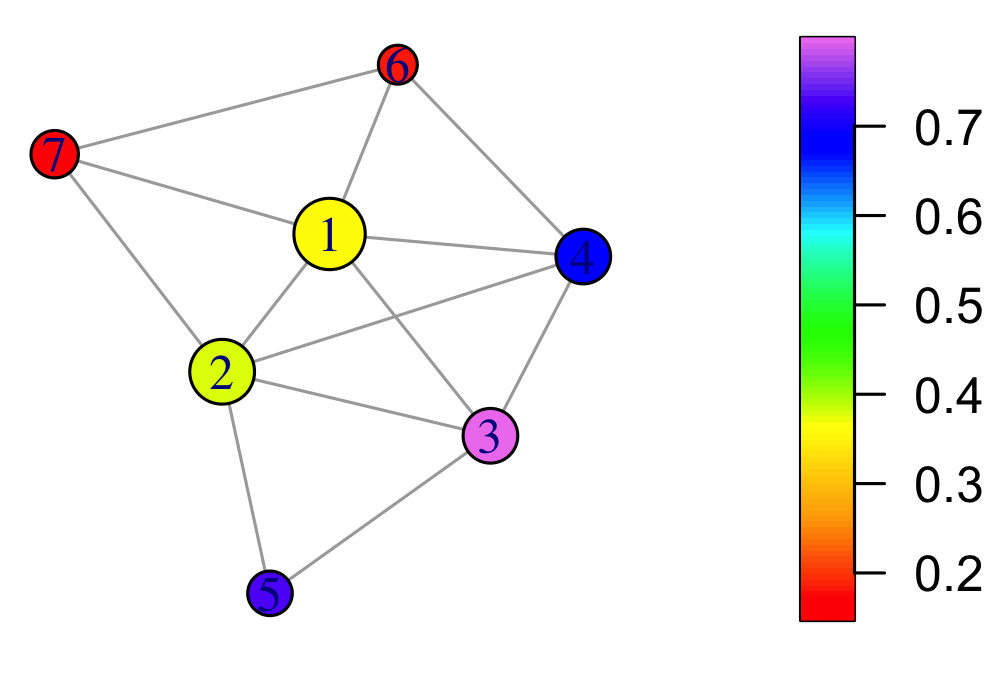}\\
			(a) Colored by $X_1$ & (b) Colored by $X_2$\\
		\end{tabular}
	\end{center}
	\raggedright
	\footnotesize{Notes: TDABM plot of a bivariate dataset $X$ in which $X_1$ and $X_2$ are drawn independently from $U(0,1)$. TDABM graph constructed with ball radius 0.4. Coloration of the two panels is according to the average value of the respective axes within each ball. TDABM implemented using the R package \texttt{BallMapper} by \cite{dlotko2019R}.}
\end{figure}

The challenge of understanding the relationship between the balls and the $X$ axis values stems from the ability of TDABM to handle multiple dimensions. In the demonstration, links with the $X$ variables is made by coloration of the balls by values of $X_1$ and $X_2$ is undertaken. Figure \ref{fig:tdabm2} has two panels in which each panel plots a different axis value. 

Figure \ref{fig:tdabm2} shows that the highest values of $X_1$ can be found to the lower left of the TDABM plot. Meanwhile, the highest values of $X_2$ are to the lower right. The bottom corner contains ball 5. Ball 5 is high for both $X_1$ and $X_2$. From Figure \ref{fig:stepsc}, it is understood that Ball 5 sits in the North East of the space. It is seen too that Ball 3 has higher values of $X_2$, but that no balls have higher $X_1$ than Ball 5. Ball 4 has low $X_1$, whilst ball 7 has low $X_1$. The lowest value of both variables in Ball 6. Again comparisons across Figures \ref{fig:stepsc} and \ref{fig:tdabm2} are clear. 

%ColorIgraphPlot(outputFromBallMapper, showVertexLabels = TRUE,
%showLegend = FALSE, minimal_ball_radius = 7,
%maximal_ball_scale = 20, maximal_color_scale = 10,
%seed_for_plotting = -1, store_in_file = "",
%default_x_image_resolution = 512, default_y_image_resolution = 512,
%number_of_colors = 100)

\section{Analyzing BallMapper Output}
\label{sec:further}

\subsection{Elements of BallMapper Output}

The \texttt{BallMapper} object that is generated through the implementation of the \texttt{BallMapper} function contains further elements which can be used to support analysis. In this section I explore the application of these additional elements for preliminary analysis of TDABM plots. The wealth of possibilities facilitated by the output go well beyond the scope of this introductory guide and readers are directed to the various papers mentioned in the introduction for inspiration. Here I offer a basic summary to help users get started analyzing their data. A list of elements in the \texttt{BallMapper} object is given in Table \ref{tab:bmo}

\begin{table}
	\begin{center}
		\caption{Elements of a \texttt{BallMapper} object}
		\label{tab:bmo}
		\begin{tabular}{|l|p{10cm}|}
			\hline
			Name & Description \\
			\hline
			vertices & These are the balls in the graph. For each vertex the number of points contained within the ball is reported \\    
			\hline        
			edges & A list of the from ball and to ball for all of the edges that are included in the TDABM plot \\
			\hline
			edges\_strength & For each edge the strength is defined as the number of points in the intersection of the to and from ball. To avoid structure created by a single point in the intersection, the edge strength can be used to manually adjust the edges list prior to passing to the plotting function. In most cases the edges\_strength element is not used.\\  
			\hline
			points\_covered\_by\_landmarks & Produces a list of the points within each ball. The points are the row numbers of each data point in the original axis dataset. In most cases the format is not useful so a separate conversion function is provided in this guide.\\ 
			\hline
			landmarks & These are the point numbers from the original axis dataset of the landmarks. Using the landmark list it is possible to merge back with the underlying dataset to learn more of the specific points that have been chosen.\\
			\hline
			coloring & A list of the coloration values for each of the balls. If you wish to provide a different value of coloring then you can do so by setting the coloring value equal to a new vector. The requirement here is that each element of the vector matches the corresponding ball in the graph.\\                    
			\hline
			coverage & A list which has an element for each data point and then a sublist within each point for the balls that contain that point. The formatting of the list is also awkward to work with so it is often easier to work from the function suggested in this guide.\\
			\hline
		\end{tabular}
	\end{center}
	\raggedright
	\footnotesize{Notes: Table provides details of the elements that are contained within a \texttt{BallMapper} object in R. TDABM is implemented using the \texttt{BallMapper} package of \cite{dlotko2019R}.}
\end{table}

\subsection{Ball Membership}

Box \ref{box:ptb} includes the \texttt{points\_to\_balls()} function which was developed by the author for the purpose of converting the output from \texttt{BallMapper} into a format that can be used to merge with the underlying dataset. The function makes use of \texttt{points\_covered\_by\_landmarks}, looping over each of the landmarks to extract the information about the points within the ball. This function can be updated for speed, but is still very quick to run in any case. The resulting dataframe, named \texttt{a1} within the function, is a long list of points for each ball starting with Ball 1. The \texttt{pt} column is the point number in the order that the original axis dataset was provided to the algorithm. 

\begin{mybox}[label=box:ptb]{R Code for Implementing TDABM}
	The points contained within each ball of the TDABM plot, \texttt{bm1}, are recovered with a user defined function:
	\begin{lstlisting}[language=R]
		points_to_balls<-function(bm1){
			a001<-length(l$landmarks)
			a1<-matrix(0,nrow=1,ncol=2)
			a1<-as.data.frame(a1)
			names(a1)<-c("pt","ball")
			for(i in 1:a001){
				a<-as.data.frame(bm1$points_covered_by_landmarks[i])
				names(a)<-"pt"
				a$ball<-i
				a1<-rbind.data.frame(a1,a)
			}
			a1<-a1[2:nrow(a1),]
			return(a1)
		}
	\end{lstlisting}
\end{mybox}

Because the output of the \texttt{points\_to\_balls} function is a long list of point numbers, the output can be merged with the underlying dataset easily. To enact the merge a column with the name \texttt{pt} is needed in the underlying dataset. The code in Box \ref{box:ptb} provides the necessary steps to get from the \texttt{BallMapper} object \texttt{bm1} to the merged dataset. The steps in Box \ref{box:ptb} include the application of the \texttt{points\_to\_balls} function.

\begin{mybox}[label=box:ptb]{Merging Balls and Underlying Data}
	We apply the function to the \texttt{bm1} object and merge with the data
	\begin{lstlisting}[language=R]
		pb1<-points_to_balls(bm1)
		names(pb1)<-c("pt","Ball")
		df1$pt<-seq(1:nrow(df1)) 
		df2<-merge(df1,pb1,by="pt")
	\end{lstlisting}
	Note here the use of \texttt{df2} recognises that points appear in multiple balls
\end{mybox}
The block of code creates a new dataframe, \texttt{df2}, which has one row for each of the points in each ball. Hence where a point appears in more than one ball, that point will appear more than once in \texttt{df2}. By keeping the merged dataset under a different name to the original data, the original data is preserved for future application. 

\subsection{Coloration}

Box \ref{box:color} provides the code for extracting the coloration of the balls in the TDABM plot and creating a data frame with the coloration values. This is useful for verifying the inference provided by the coloration in the TDABM graph itself. 

\begin{mybox}[label=box:color]{R Code for Implementing TDABM}
	The TDABM object, \texttt{bm1}, can be plotted in R using:
	\begin{lstlisting}[language=R]
		cd1<-as.data.frame(bm1$color)
		cd1$Ball<-seq(1:nrow(cd1))
		names(cd1)<-c("TDABM_Coloration","Ball")
	\end{lstlisting}
\end{mybox}

\begin{figure}
	\begin{center}
		\caption{TDABM Plot}
		\label{fig:sdy}
		\includegraphics[width=12cm]{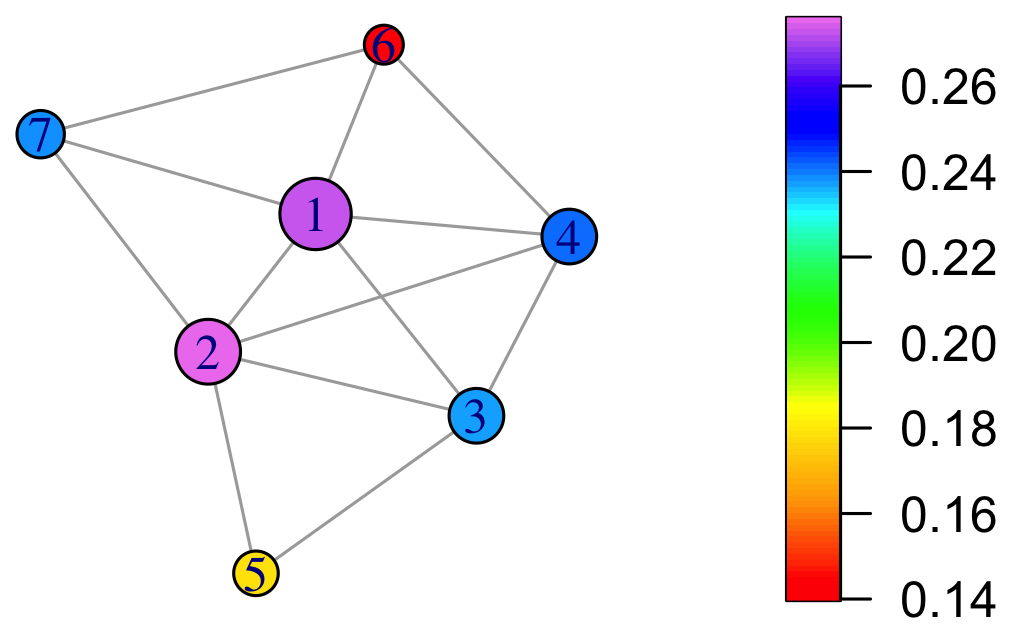}
	\end{center}
	\raggedright
	\footnotesize{Notes: TDABM plot of a bivariate dataset $X$ in which $X_1$ and $X_2$ are drawn independently from $U(0,1)$. TDABM graph constructed with ball radius 0.4. The generation process for the outcome $Y$ has $y_i = x_{i1} + x_{i2}$. Coloration is according to the standard deviation of $Y$ within each ball. TDABM implemented using the R package \texttt{BallMapper} by \cite{dlotko2019R}.}
\end{figure}

The coloration element can also be used in reverse to generate a new coloration variable for the plot. Using the merged dataset, summaries for each ball can be generated by using \texttt{dplyr} to group on the ball number. If a new coloration dataset emerges then the coloring element can be udpated by assignment in R. The code in Box \ref{box:color2} shows the construction of the standard deviation of $Y$ for each data point and then the subsequent assignment of the coloration to the \texttt{bm1} object. Figure \ref{fig:sdy} gives the updated TDABM plot.

\begin{mybox}[label=box:color2]{R Code for Generating a new Coloration}
	First summarise the extended dataframe by grouping on the \texttt{Ball} variable:
	\begin{lstlisting}[language=R]
		df2g<-group_by(df2,Ball)
		df2s<-summarise(df2g,sdy = sd(Y), .groups="drop")
		df2s<-as.data.frame(df2s)
		names(df2s)<-c("Ball","SDY")
	\end{lstlisting}
	Next update the coloring variable
	\begin{lstlisting}[language=R]
		bm1$coloring<-df2s$SDY
	\end{lstlisting}
	The plot of \texttt{bm1} can then be generated in the usual way.
\end{mybox}

Figure \ref{fig:sdy} reveals that there is a large variation in $Y$ within each ball. The result is unsurprising since a large radius is being used and the values of $Y$ are tied to the values of both axis variables. The smallest standard deviation appears in the smaller balls at the edges of the plot. Balls 5 and 6 both have lower standard deviations. Here the summary function is purely for illustration. In other contexts, the extent of within ball variation in outcome may be used as a guide on radius selection, or as a way to understand the ability of the variables in $X$ to explain the variation in $Y$. Note that TDABM is only showing the extent of variation of $Y$ conditional on the $X$ within the ball, there is no causality implied. Adding additional coloration variables requires updating the code in Box \ref{box:color2}.

\subsection{Updating the Plot Function}

Finally, this guide considers the ways in which the \texttt{ColorIgraphPlot} function may be updated to provide more meaningful plots. The inputs to the \texttt{ColorIgraphPlot} function are provided in Table \ref{tab:cig}.

\begin{table}
	\begin{center}
		\caption{Inputs to ColorIgraphPlot}
		\label{tab:cig}
		\begin{tabular}{|l|l|p{8cm}|}
			\hline
			Input Element & Default & Description \\
			\texttt{outputFromBallMapper} &  & \texttt{BallMapper} object as generated by the \texttt{BallMapper} function \\
			\hline
			\texttt{showVertexLabels} & TRUE & A TRUE/FALSE indicator that states whether the numbers should be added to the balls. Numbers are useful for discussing the balls, but may distract in some applications. This option can be useful when plotting axes and having one plot with ball numbers against which the others can be directly compared.\\
			\hline
			\texttt{showLegend} & FALSE & This will add a list of balls to the plot. Typically, only the color bar is needed for interpreting the TDABM graph\\
			\hline
			\texttt{minimal\_ball\_radius } & 7& Controls the size of the smallest ball. The size should be sufficient to include the label of the ball. Hence the minimum is defaulted to 7.\\
			\hline
			\texttt{maximal\_ball\_scale } & 20& Controls the maximum size of a ball in the TDABM plot. This value can be made larger, but it must be remembered that large balls will start to dominate the space.\\
			\hline 
			\texttt{maximal\_color\_scale } & 10&\\
			\hline
			\texttt{seed\_for\_plotting } & -1 & Because TDABM creates abstract visualisations, the process through which the balls are represented in two dimensions applies a spring algorithm. Setting the seed ensures that reproducability of the spring algorithm. Hence set the seed to have a consistent TDABM outcome.\\
			\hline
			\texttt{ store\_in\_file = ""}&& Allows the specification of a filename to store the resulting TDABM plot. In this guide the code is provided to open a new device and then plot the TDABM graph. You may prefer to set the inputs here.\\
			\hline
			\texttt{default\_x\_image\_resolution = 512} & 512 & The resolution used on the horizontal direction defines the quality of the TDABM plot when it appears in documents. Higher resolutions improve clarity but also generate larger files. Use of little \texttt{x} here is because of the R function thinking of the horizontal axis.\\
			\hline
			\texttt{default\_y\_image\_resolution = 512} & 512 & The resolution used on the vertical axis further defines the quality of the TDABM plot when it appears in documents. Higher resolutions improve clarity but also generate larger files. Use of little \texttt{y} here is because of the R function thinking of the vertical axis.\\
			\hline
			\texttt{number\_of\_colors }&100& The number of colors is a parameter of the colorbar in R. Increasing the number allows for more subtle variations between the colors of balls.\\
			\hline
		\end{tabular}
	\end{center}
		\raggedright
	\footnotesize{Notes: Table provides details of the inputs that are are provided to the \texttt{ColorIgraphPlot} function as implemented in the \texttt{BallMapper} package of \cite{dlotko2019R}.}
\end{table}

Table \ref{tab:cig} demonstrates the flexibility of the plotting of TDABM graphs. It is possible to dig further into the options by going into the R function itself and making adjustments. For most purposes, the default inputs to the function are sufficient. Elements such as turning the ball labels off can be used to help readability of plots. The numbers are for discussing the balls, they do not have any value relevant function in the analysis. Hence, provided there is at least one plot with numbers it is possible to have other versions of the same graph without the numbers. The options to specify graphic parameters are also useful, but here the code is used to generate the plot as a new graphic device in R. The saving of plots is done using separate blocks of code.

\section{Summary}
\label{sec:summary}

This guide is an introduction to the \texttt{BallMapper} function from the \texttt{BallMapper} package in R \citep{dlotko2019R}. Using a simple bivariate dataset, it is shown how the BallMapper algorithm of \cite{dlotko2019ball} is implemented. It is then shown how the R package interprets the selection of landmarks and how the resulting difference in TDABM graph appearance results. The dataset is artificial and as such there is limited interpretation to be gained from the balls that are generated. However, as demonstrated in the literature, the plots do have meaning once the data is from the real-world \citep[amongst others]{rudkin2023economic,otway2024shape,rudkin2024topology}. Readers are encouraged to apply TDABM to their own datasets as part of appreciating the structures that exist within their data. 

There are many alternative means to visualize multidimensional data. However, the simplicity of the TDABM approach gives a ready interpretability. Users may compare the results of the TDABM against alternatives on their data. In the literature, it is the stability of the TDABM graphs, the simplicity of having a single parameter, and the neat interpretation of the similarity of data points within each ball, that motivates adoption. The method also begs comparison with clustering algorithms, but it must be stressed that TDABM is not a clustering algorithm. The grouping of observations into balls is simply to provide a reduction in the number of points for visualizing. Unlike almost all clustering algorithms, the balls of TDABM are all equal in size within the characteristic variable space. Comparisons between balls and clusters is also commonly incorporated into papers applying TDABM. The toolkit built upon TDABM is constantly evolving. Functionality is available in both Python and R. This guide is based upon R, but the principles can be readily matched to Python. 

The need to visualize data is emphasised by \cite{matejka2017same}, whilst the need to evaluate statistical models with visualization is shown in \cite{anscombe1973graphs}. This guide shows how TDABM can produce data visualizations, and how the algorithm allows coloration. As  \cite{qiu2020refining}, \cite{dlotko2021financial}, \cite{rudkin2024return} and others discuss, TDABM offers a means to evaluate models. TDABM has a contribution at the start of the modelling process to understand data structure, and at the end to plot model performance. In all analysis conducted to date, the application of TDABM has revealed new information in the data. Future research will continue to expand upon the role of TDABM in statistical analysis and as a visualization tool. This guide shows how the \texttt{BallMapper} function in R \citep{dlotko2019R} provides foundation to that research agenda.

\bibliography{tdabmpres}
\bibliographystyle{apalike}

\end{document}